\documentclass{article}
\usepackage{arxiv}
\usepackage[utf8]{inputenc} 
\usepackage[T1]{fontenc}    
\usepackage{hyperref}       
\usepackage{url}            
\usepackage{booktabs}       
\usepackage{amsfonts}       
\usepackage[ruled, linesnumbered]{algorithm2e}
\usepackage{amssymb}
\usepackage{amsmath}
\usepackage{graphicx}
\usepackage{psfrag}
\usepackage{epsfig}
\usepackage{subfig}
\usepackage{paralist}
\usepackage{mdwlist}
\usepackage{multirow}
\usepackage{comment}
\usepackage{pstricks}
\usepackage{psfrag}
\usepackage{color, colortbl}

\newtheorem{example}{Example}

\newcommand{\qed}{\hfill $\Box$}

\title{Early-stage Resource Estimation from\\Functional Reliability
Specifications in\\Embedded Cyber-Physical Systems}

\author{
  Ginju V. George\\
  Indian Space Research Organization (ISRO),\\
  Vikram Sarabhai Space Centre (VSSC)\\
  Trivandrum, Kerala, India -- 695022\\
  \texttt{ginjuvgeorge@gmail.com}\\
   \And
  Aritra Hazra, Pallab Dasgupta and Partha Pratim Chakrabarti \\
  Department of Computer Science and Engineering\\
  Indian Institute of Technology Kharagpur\\
  Paschim Medinipur, West Bengal, India -- 721302\\
  \texttt{\{aritrah, pallab, ppchak\}@cse.iitkgp.ac.in} \\
}

\begin{document}

\maketitle

\begin{abstract}
Reliability and fault tolerance are critical attributes of embedded
cyber-physical systems that require a high safety-integrity level. For such
systems, the use of formal functional safety specifications has been strongly
advocated in most industrial safety standards, but reliability and fault
tolerance have traditionally been treated as platform issues. We believe that
addressing reliability and fault tolerance at the functional safety level widens
the scope for resource optimization, targeting those functionalities that are
safety-critical, rather than the entire platform. Moreover, for software based 
control functionalities, temporal redundancies have become just as important as
replication of physical resources, and such redundancies can be modeled at the
functional specification level. The ability to formally model functional
reliability at a specification level enables early estimation of physical
resources and computation bandwidth requirements. In this paper we propose, for
the first time, a resource estimation methodology from a formal functional
safety specification augmented by reliability annotations. The proposed
reliability specification is overlaid on the safety-critical functional
specification and our methodology extracts a constraint satisfaction problem
for determining the optimal set of resources for meeting the reliability target
for the safety-critical behaviors. We use SMT (Satisfiability Modulo Theories) /
ILP (Integer Linear Programming) solvers at the back end to solve the
optimization problem, and demonstrate the feasibility of our methodology on a
Satellite Launch Vehicle Navigation, Guidance and Control (NGC) System.
\end{abstract}

\keywords{Formal Specification, Resource Allocation, Processor Minimization}


\maketitle

\section{Introduction} \label{sec: introduction}
With increasing dependence of safety-critical systems on software and embedded
computing, the notions of reliability and fault tolerance have undergone a
paradigm shift. While the traditional approach for enhancing reliability in
electro-mechanical systems is based on replication of physical
components~\cite{KH1980TR, KS1996TC, LCE1989FTCS, SK1994TC, AK1984C,
LA1978FTCS}, safety-critical embedded cyber-physical systems (CPS) must also
consider reliability of software executions, packet drops in networks, priority
inversion in scheduling of control tasks on shared processors, and many other
factors which are attributed to the computational platform which executes the
{\em cyber} part of the system~\cite{CACGWZ2016DT, L2008ISORC}.

In the domain of cyber-physical systems, reliability targets of embedded control
can be met through fault-tolerant design of the underlying platform and its
structural components or through fault-tolerant deployment of the control
software on the underlying platform. The former looks at the necessary resource
requirements for providing reliable service at the platform level without
specific focus on individual control components~\cite{BFMSPP2003CASES}. For 
example, while designing the CAN matrix for an automobile, we choose the 
topology and number of CAN segments depending on the overall loading of the 
buses and the reliability concerns thereof. On the other hand, fault-tolerant 
deployment of control focuses on individual control tasks~\cite{GMBSC2014ISIC}. 
For example, it includes replication of software tasks on multiple 
cores~\cite{PCS2008TCAD} and sometimes on the same core (temporal 
redundancy)~\cite{GRW1988TR, KS1993TR, K1999RTCSA, KK2004JRTS}, or provides
fault-tolerant scheduling frameworks that involve cloning of a sequence of tasks 
for layered task graphs~\cite{DDD1998JPDC, PLW1996TPDS}. Fault-tolerant 
deployment also influences the choice of the platform and its components, but 
with respect to the reliability requirements for individual functional 
components.

The design cycle of cyber-physical systems begin with the preparation of a
specification of the functionalities of the system. The use of formal
specifications for defining the functional safety properties of such systems has
been widely recommended in most industrial standards, including
aeronautics (DO-178C), automotive (ISO 26262), industrial process automation
(IEC 61508), nuclear (IEC 60880), railway (EN 50128) and space (ECSS-Q-ST-80C),
specifically for those functionalities that require a high safety-integrity
level. Although reliability and fault tolerance are as important attributes of
the system design as functional correctness~\cite{BCCZ1999LNCS, CBRZ2001FMSD,
CGP2000BOOK, D2006BOOK, L2005BOOK, M1992PHDTHESIS}, or performance attributes
such as timing~\cite{AD1994TCS, AIKY1992CAV, DDR2010DATE, DRD2014FAC,
M2014LNCS}, power~\cite{HGDP2013TVLSI, HMDPHBM2013TCAD} and
security~\cite{KRH2017DAC}, formal specification of reliability, especially with
respect to the critical functional safety properties has so far received very
little attention. This is largely due to the perception that reliability and
fault tolerance need to be addressed at the platform level, not at the
functional level~\cite{J1989BOOK, SS1992BOOK}. On the other hand, we believe
that with the increasing cost of electronics and software in cyber-physical
systems, investment on reliability needs to be prioritized on the basis of the
safety-criticality of the system's functions.

This paper presents, for the first time, a methodology for overlaying formal
functional safety specifications with reliability annotations and estimating the
resource requirement from such extended specifications. A formal specification
enables the designer to lay out the strategy for redundant computations and/or
actuations~\cite{HDC2016JAL} and obtain a formal reliability guarantee for the
strategy using the proposed method of analysis. Moreover since this is done at
the specification level, our methodology provides early estimates of the
resource requirements, thereby facilitating design space exploration, where the
tradeoff between reliability and resource requirements can be studied using the
proposed methodology until an acceptable balance is achieved.

In particular, our previous work~\cite{HDC2016JAL} explores various reliable
strategies enabled by the given reliability specifications and converges on a
strategy that maximizes the reliability -- thereby also pointing out whether the
specified reliability targets are attainable for every functionality of the
design. In our level of abstraction, an {\em action} represents a discrete
control event which is enabled by a logically defined pre-condition ({\em
sense}) and achieves a logically specified consequent ({\em outcome}) when
executed successfully on the underlying computational platform. We use the term
{\em action-strategy} to denote possible sequences of actions that lead to some 
desired outcome. An action-strategy is {\em admissible} with respect to a 
reliability target, if the sequence and timing of actions involved in that 
strategy guarantees the desired outcome with the specified reliability 
guarantee. It may be noted that, more than one admissible action-strategies may 
exist for a functionality and every participating action (at any time instant) 
in an admissible strategy requires one processor (computing resource) to 
execute. 

In the early stage of design of a complex, multi-component embedded control 
system, the designer must plan the sense-control-actuation steps for each control
loop along with the corresponding timing constraints. The formal safety 
specification at this stage defines the necessary outcomes of the closed loop 
system as a timed function of the sensed input events. For example, we may be
given a safety property of the form:
\[ {\tt 
	sense\_A \rightarrow \text{\#\#}[1:10]\ outcome\_B
	} \]
where {\tt sense\_A} is a sense event and {\tt outcome\_B} is the desired 
outcome which must happen within the time interval {\tt [1:10]} following the
sense event. Now in order to ensure that {\tt outcome\_B} does actually happen
in that window of time, the control system must perform appropriate actions 
(called actuations).

Suppose we have an action, called {\tt action\_C} which can cause {\tt outcome\_B},
but the causality is not fully guaranteed. Suppose:
\[ Prob({\tt outcome\_B}\ |\ {\tt action\_C}) = 0.8 \]
which means that {\tt action\_C}, if used, can cause {\tt outcome\_B} 80\% of the 
time. Now suppose our reliability target for the above safety property is $0.95$.
If the sense event, {\tt sense\_A}, triggers one execution of the action, 
{\tt action\_C}, then the reliability with which {\tt outcome\_B} is guaranteed
is only $0.8$. Therefore, in order to achieve the target of $0.95$, we must be
prepared to repeat {\tt action\_C} more than once, either on different resources, or 
on the same resource but at different times. Moreover, all executions must be completed
within the specified window of time, namely {\tt [1:10]}, following the event 
{\tt sense\_A}.

In a complex system, there can be many safety properties and therefore, many 
different actions may have to be repeated spatially and/or temporally to achieve the 
desired reliability. We provide a framework, where the system designer can overlay
a chosen pattern of redundancy over the temporal fabric of the safety specification.
For example, consider the following specification:
\[ {\tt
	sense\_A \rightarrow \text{\#\#}[2:5]\ action\_C[\sim 2]\
	\text{\#\#}[1:2]\ action\_C
	} \]
The above specification specifies that following a {\tt sense\_A} event, the action,
{\tt action\_C}, will be executed in parallel at some time in the window {\tt [2:5]},
and will be again executed after $[1:2]$ units of time. It be noted that all three 
executions of {\tt action\_C} are constrained by time windows, but they can be 
executed in more than one way within the specified windows. A formal specification
framework like this has two immediate advantages:
\begin{enumerate}

\item We can verify whether the specified redundancy in the executions of the actions
	achieves the target reliability of the safety properties.

\item Each possible schedules of the actions provides us an early estimate of the resource
	requirements. We can also find the schedule which has the optimal resource 
	requirement. In a complex system, with many actions and replications, this is a
	non-trivial task and computationally beyond manual capacity.

\end{enumerate}
It may also be noted that two admissible action-strategies with respect to two different 
reliability specifications may share some common actions. Given a set of admissible 
action-strategies with respect to various sensed-events, it is necessary to predict the 
amount of resources (processors) required in order to execute all of them in a 
worst-case scenario when all sensed-events, take place together. This estimation is 
non-trivial due to the sharing of actions and the redundancy provisions present in the 
strategies, and hence the choice of an appropriate admissible action-strategy with 
respect to every sensing is a key to reduce the number of resources. This paper 
presents a novel technique for selecting and grouping the set of admissible 
action-strategies attributed from various specifications with respect to different 
sensed-events, so that the number of resources needed in the worst case for 
executing the actions is minimized.

With the rapid growth in the number of features supported in modern appliances, the
computing paradigm is shifting from a federated architecture where each feature runs
on a dedicated processor to an integrated architecture where Electronic Control Units
(ECUs) are shared among multiple features. While integrated architectures are able to
reduce the costs of electronics and networking, the scheduling 
and fault tolerant deployment of control tasks has become more complex. We believe
that having an early estimate of the resource requirements, as in the proposed 
approach, can help in taking rational decisions on the tradeoff between reliability and
resource costs.

Our early work, presented in~\cite{HDC2016JAL}, introduces the formal framework for 
reliability specification. The problem of estimating resource requirements from such 
specifications is treated in this paper for the first time. The main contributions of this
paper are as follows:
\begin{itemize}
 
\item We extract all possible admissible reliability strategies from a given
	reliability specification, that meets the desired reliability target 
	for a functional specification.

 \item We formulate the resource optimization problem in terms of a
	constraint-satisfaction problem which can be solved using SMT / ILP
	solvers.

 \item We have studied the proposed technique over test-cases from automotive
	domain and also shown the scalability of our approach over several
	experiments.

 \item We illustrate the practicality of our proposed technique using a
	case-study (over Satellite Launch Vehicle Navigation, Guidance and 
	Control (NGC) System) from avionics domain.

\end{itemize}
It is important to note that resource estimation is intricately related to the code volumes
associated with the control tasks and their worst case execution times (WCET) on the
chosen ECUs. In practice, the design of embedded systems is a process that goes
through multiple generations of a product, and statistical knowledge about the 
execution times and the probability of their success in terms of providing the desired
outcome are available from legacy data. Therefore early estimation of the tradeoff
between functional reliability and resource requirements is feasible in practice,
though not in established practice today.

The paper is organized as follows. Section~\ref{sec: background} illustrates the
preliminary concepts and the earlier works carried out to set up the premise
for this work. In Section~\ref{sec: rsrc_est}, we present the formal problem
definition and methods to determine the optimal number of resources. The 
experimental results showing the efficacy of our proposed framework is given in 
Section~\ref{sec: exp_results}. In Section~\ref{sec: case-study}, we demonstrate 
our proposed methodology using a practical case-study.
Finally, Section~\ref{sec: conclusion} concludes the paper with a brief 
discussion on the ramifications of the proposed approach.

\section{Background Work and Preliminaries} \label{sec: background}
An embedded controller of a CPS comprises of three primary modules, namely {\em
Sensor}, {\em Controller} and {\em Actuator} -- which interacts with the {\em
Plant} (or the environment) to perform desired behaviors. The {\em sensor} is 
responsible for sensing input scenarios from the plant (or the environment). The 
{\em controller} performs the control decision based on the sensed input. The 
{\em actuator} delivers actuation signals to the plant (or the environment). 
Some of the activities in CPS control may be implemented using software and the 
rest may be controlled via hardware. For example, sensing a scenario where the 
brake should be applied automatically, the computation of appropriate 
brake-pressure may be carried out using a software program. However, maintaining 
the same break-pressure for a certain amount of time is performed by a 
hardware-control. The resultant outcome while applying such actions is the 
application of the wheel-brakes and thereby reduction in the vehicle speed. We 
present a formal model for embedded CPS and such requirements in the following.

\begin{figure}[htb]
\begin{center}
\psfrag{EI}{$\mathcal{E}_I$}
\psfrag{EA}{$\mathcal{E}_A$}
\psfrag{EO}{$\mathcal{E}_O$}
\includegraphics[scale=0.5]{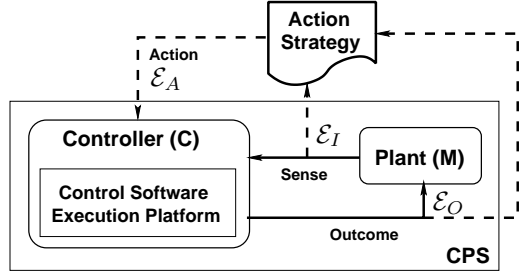}
\end{center}
\caption{Schematic Representation of the Embedded CPS Framework}
\label{fig: system_model}
\end{figure}

Figure~\ref{fig: system_model} provides a schematic representation of the
embedded CPS framework. Here, the events that are responsible for sensing
inputs, actions (actuations) and outcome activities are termed as {\em
sensed-events} ($\mathcal{E}_I$), {\em action-events} ($\mathcal{E}_A$) and
{\em outcome-events} ($\mathcal{E}_O$), respectively.  While interacting with
the plant model, the controller receives the sensed-events and compute
appropriate actions/actuations to produce the desired outcome-events. The
outcome of an action may take place after many control cycles and may also be
durable over a period of time. Typically, the {\em control software execution
platform} is responsible to compute appropriate actuations for the desired
outcomes being enabled.

The reliability of the CPS is governed by the fault-free execution from sensed
to the outcome events. This intuitively means, after sensing an input scenario, 
how reliably an action (actuations) can be performed once the controller 
decision is undertaken properly. In this model, the reliability of the outcome 
activities are attributed from the unreliable computational platform where the 
action events are scheduled, computed and applied. To meet the  desired 
reliability of the outcome behavior, the fine-grain control incorporates various 
physical redundancy attributes in the scheduled actions inside the control 
execution platform. It may be noted that the unreliability lies primarily in the 
execution of action-events (as per our model), though there can also be failures 
in the sensing and application of outcome activities or imperfections in message 
communications. However, this framework can be suitably extended to handle such 
uncertainties in other components of the embedded CPS framework.

From an abstract point of view, the activities of a CPS can be visualized in 
terms of sequence of events. The early-stage functional specification narrates 
the outcome-event based on a sensed scenario which is to be met with specific 
reliability targets. Such specifications describe the correctness requirements 
for a CPS. In addition to this, the meta-level {\em action-strategy} specifies
the supervisory control orchestrated by the control software where the 
redundancies in actions are described for a sensed scenario. Such requirements 
are termed as CPS reliability specifications~\cite{HDC2016JAL}.

Now, let us revisit the formal model for such embedded CPS specifications as
introduced in~\cite{HDC2016JAL} using an elementary correctness requirement,
$\mathcal{P}_S$, for the subsystem, $S$, with a sense-event, $E_I \in
\mathcal{E}_I$, and an outcome-event, $E_O \in \mathcal{E}_O$ as follows:
\[
\mathcal{P}_S:\ \ {\tt ( E_I \rightarrow \text{\#\#} [a:b]\ E_O} )
\]
It means that, whenever $E_I$ is sensed then the outcome-event $E_O$ is
asserted within next ${\tt a}$ to ${\tt b}$ time-units\footnote{The formal
expressions of the properties have syntactic (and semantic) similarity with
SystemVerilog Assertions (SVA)~\cite{system_verilog}.}. Suppose, $E_A \in
\mathcal{E}_A$ is the action-event responsible for producing the outcome
$E_O$. If the probability of success for $E_O$ given the successful execution
of $E_A$ is expressed as, $R_{E_O} = Prob(E_O\ |\ E_A)$, then the current
reliability ($R_{\mathcal{P}_S}$) of the property, $\mathcal{P}_S$, is also 
$R_{E_O}$. Since we have $R_{\mathcal{P}_S} = Prob(E_O\ |\ E_A)\ast Prob(E_A\ 
|\ E_I)\ast Prob(E_I)$, so $R_{\mathcal{P}_S} = R_{E_O}$ happens under the 
assumption of a perfect sense and action scenario\footnote{We assume that the 
control system senses perfectly (without any false positives or false 
negatives) and also generates appropriate actions based on the sensed-event.}, 
where both $Prob(E_A\ |\ E_I) = 1$ and $Prob(E_I) = 1$.

\begin{figure*}[htb]
\begin{center}
\subfloat[Spatial Redundancy Model]
 {
 \psfrag{C1}{\small $C_1$}
 \psfrag{C2}{\small $C_2$}
 \psfrag{Cn}{\small $C_n$}
 \psfrag{M1}{\small $M_1$}
 \psfrag{M2}{\small $M_2$}
 \psfrag{Mn}{\small $M_n$}
 \psfrag{EI}{\small $E_I$}
 \psfrag{EC1(t)}{\small $E_{A_1}(t)$}
 \psfrag{EC2(t)}{\small $E_{A_2}(t)$}
 \psfrag{ECn(t)}{\small $E_{A_n}(t)$}
 \psfrag{EO}{\small $E_O$}
 \psfrag{V}{\small $V$}
 \includegraphics[scale=0.5]{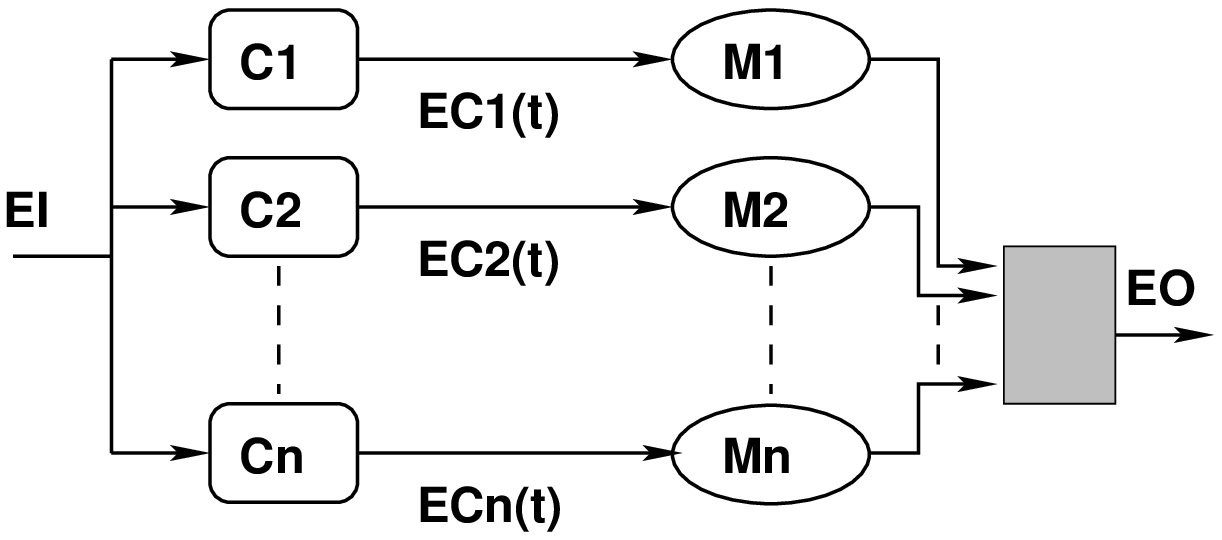}
 }
\quad
\subfloat[Temporal Redundancy Model]
 {
 \psfrag{t1}{\small $t_1$}
 \psfrag{t2}{\small $t_2$}
 \psfrag{tm}{\small $t_m$}
 \psfrag{C}{\small $C$}
 \psfrag{Mt1}{\footnotesize $M(t_1)$}
 \psfrag{Mt2}{\footnotesize $M(t_2)$}
 \psfrag{Mtm}{\footnotesize $M(t_m)$}
 \psfrag{EI}{\small $E_I$}
 \psfrag{ECt1}{\small $E_A(t_1)$}
 \psfrag{ECt2}{\small $E_A(t_2)$}
 \psfrag{ECtm}{\small $E_A(t_m)$}
 \psfrag{EO}{\small $E_O$}
 \psfrag{V}{$V$}
 \includegraphics[scale=0.5]{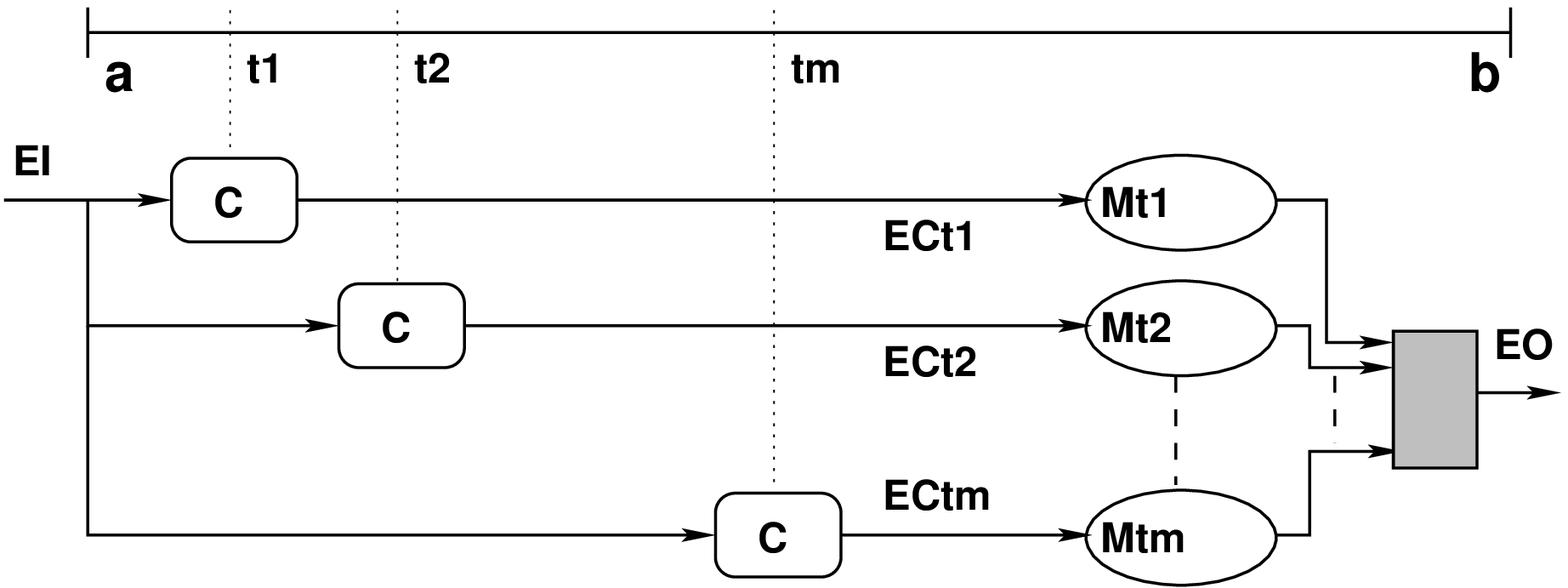}
 }
\end{center}
\caption{Redundancy Models of an Embedded CPS Control (adopted
from~\cite{HDC2016JAL})}
\label{fig: red_model}
\end{figure*}

At the design-level, the controller can leverage spatial and temporal redundancy
provisions to improve the reliability of the outcome-events and thereby enhance
the functional reliability of the design specifications. Figure~\ref{fig: 
red_model}(a) and Figure~\ref{fig: red_model}(b) show the schematic models 
where the spatial and temporal redundancy provisions are incorporated at the 
controller level. In the spacial redundancy model for a CPS control system, 
$S$, there are $n$ parallel instances of the controller, namely $C_1, C_2, 
\ldots, C_n$ and $n$ computational counterparts, $M_1, M_2, \ldots, M_n$.  The 
controllers $C_i, i \in [1,n]$ sense the same events ($E_I \in \mathcal{E}_I$) 
and produce the action-events $E_{A_1}(t), E_{A_2}(t), \ldots, E_{A_n}(t)$ 
($\forall i \in [1,n],\ E_{A_i}(t) \in \mathcal{E}_A$), respectively, at time 
$t$. Each action-event $E_{A_i}$ goes into a possibly unreliable computational 
counterpart, $M_i$, respectively. On the other hand, in the temporal redundancy 
model, $m$ number of re-executions are made by a single controller within the 
time-window from ${\tt a}$ to ${\tt b}$. At time $t_i$ ($\forall i \in [1,m],\ a 
\leq t_i \leq b$), the controller, $C$, senses same input-event, $E_I \in 
\mathcal{E}_I$, and produces the action-event, $E_A(t_i) \in \mathcal{E}_A$. 
This action-event goes into the possibly unreliable computational counterparts, 
$M(t_i)\ (i \in [1,n])$, and produces an output-event at $t_i$. Finally, the 
outcome-event, $E_O \in \mathcal{E}_O$, produces successful (correct) outcome if 
one of the outputs of $M_i$ (or $M(t_i)$) produces correct result for both 
these cases\footnote{We assume a fail-silent model here, where failure in any 
outcome implies discontinuation of that outcome -- thereby it will not 
interfere with other correct outcomes and can possibly be distinguished from 
these correct outcomes.}.

Such design-level redundancy provisions can be captured from the functional
specification-level as well. For example, we can formally represent the spatial
and temporal redundancies (as shown in Figure~\ref{fig: red_model}) using the
properties, $\mathcal{P}_S^{Srel}$ and $\mathcal{P}_S^{Trel}$, respectively,
as follows:
\[
\mathcal{P}_S^{Srel} :\ \ {\tt ( E_I \rightarrow \text{\#\#} [a:b]\ (E_A
\text{[}\sim n\text{]}) )} \ \ \ \ \ \text{and} \ \ \ \ \ 
\mathcal{P}_S^{Trel} :\ \ {\tt ( E_I \rightarrow \text{\#\#} [a:b]\ (E_A
\text{[}=m\text{]}) )}
\]
Here, $\mathcal{P}_S^{Srel}$ asserts that, whenever $E_I$ is sensed then the
action-event $E_A$ is asserted parallelly in $n$ execution units (denoted by
the ${\tt [\sim n]}$ construct) within next ${\tt a}$ to ${\tt b}$ time-units.
Similarly, $\mathcal{P}_S^{Trel}$ asserts that, whenever $E_I$ is sensed then
the action-event $E_A$ is asserted $m$ times by multiple executions (denoted
by the ${\tt [=m]}$ construct) within next ${\tt a}$ to ${\tt b}$ time-units. At
this current setup, the reliability of the outcomes are also increased and these
are represented as, $R_{E_O}^{Srel} = 1 - (1 - R_{E_O})^n$ and
$R_{E_O}^{Trel} = 1 - (1 - R_{E_O})^m$, respectively.

Let us present an example to explain such functional correctness and reliability
specifications.

{\small
\begin{example} \label{ex: acc_rel}
{\em Adaptive Cruise Control (ACC) supports features to -- (a) maintain a 
minimum following interval to a lead vehicle in the same lane and (b) control 
the vehicle speed whenever any lead obstacle is present. Let us consider the
functional requirements of ACC, which senses the proximity of any lead vehicle
and a leading obstacle by the sense-events, ${\tt lead\_gap}$ and ${\tt
lead\_obs}$, respectively. The required action-events issued by the ACC
controller for reducing throttle by $10\%$ and applying proper pressure in
wheel-brakes are denoted as, ${\tt act1}$ and ${\tt act2}$, respectively. The
corresponding outcome-events are given as, ${\tt thrt\_adj}$ and ${\tt
brk\_adj}$. It may be noted that a single time-unit delay is considered to be
of $50$ {\tt ms} in all the functional specifications mentioned below.

Now, consider the following two functional correctness requirements of ACC as 
follows:
\begin{enumerate}[(a)]
 \item ${\tt ACC\_C1}$:
	{\em As soon as a lead obstacle is sensed, then within a total of $200$
	{\tt ms}, the throttle is adjusted (reduced by $10\%$) followed by the
	application of wheel brakes after $50$-$100$ {\tt ms}}. Formally, this
	property is expressed as:
	\[ {\tt 
	ACC\_C1:\ lead\_obs \rightarrow \text{\#\#}[1:2]\ thrt\_adj\
	\text{\#\#}[1:2]\ brk\_adj
	} \]

 \item ${\tt ACC\_C2}$:
	{\em Whenever a lead vehicle is sensed in a close proximity, then within
	a total of $300$ {\tt ms}, the throttle is adjusted (reduced by $10\%$)
	followed by the application of wheel brakes after $50$-$150$ {\tt ms}}.
	Formally, this property is expressed as:
	\[ {\tt 
	ACC\_C2:\ lead\_gap \rightarrow \text{\#\#}[1:3]\ thrt\_adj\
	\text{\#\#}[1:3]\ brk\_adj
	} \]
\end{enumerate}

Suppose, the desired reliability of the given correctness requirements be $0.95$
and $0.98$ for ${\tt ACC\_C1}$ and ${\tt ACC\_C2}$, respectively. Here, the
actions ${\tt act1}$ and ${\tt act2}$ are responsible for producing the
outcome-events ${\tt thrt\_adj}$ and ${\tt brk\_adj}$, respectively. Suppose,
the outcome-events are unreliable and their reliability values are given as:
\[
R_{\tt thrt\_adj} = Prob({\tt thrt\_adj}\ |\ {\tt act1}) = 0.8\ \text{ and }\
R_{\tt brk\_adj} = Prob({\tt brk\_adj}\ |\ {\tt act2}) = 0.9.
\]

\noindent Since the outcomes are unreliable, the ACC must issue the
action-events with appropriate redundancy in order to meet the desired
reliability target, as given by the following reliability specifications.
\begin{enumerate}[(a)]
 \item ${\tt ACC\_R1}$:
	{\em As soon as a lead obstacle is sensed, then the throttle-reduction
	action-event is scheduled in two processors parallelly within $50-100$
	{\tt ms}, followed by the brake-apply action-event which is also
	scheduled in two processors parallelly within next $50-100$ {\tt ms}}.
	Formally, this property is expressed as:
	\[ {\tt
	ACC\_R1:\ lead\_obs \rightarrow \text{\#\#}[1:2]\ act1[\sim 2]\
	\text{\#\#}[1:2]\ act2[\sim 2]
	} \]

 \item ${\tt ACC\_R2}$:
	{\em Whenever a lead vehicle is sensed in a close proximity, then the
	overall action of $10\%$ throttle-reduction applied successively twice,
	followed by brake-application in the next time-unit is re-executed twice
	within an overall time limit of $300$ {\tt ms}}. Formally, this property
	is expressed as:
	\[ {\tt ACC\_R2:\ lead\_gap \rightarrow\ \text{\#\#}[1:3]\ (act1[*2]\
	\text{\#\#}1\ act2)[=2]
	} \]
\end{enumerate}

Now, given the reliability specifications, ${\tt ACC\_R1}$ and ${\tt ACC\_R2}$
and assuming that the sensed-events (${\tt lead\_obs}$ and ${\tt lead\_gap}$) are
present in Cycle-0, the reliability for the correctness specifications, ${\tt
ACC\_C1}$ and ${\tt ACC\_C2}$ are calculated in Table~\ref{table: acc_r1} and
Table~\ref{table: acc_r2}, respectively, with respect to each action/control
strategy\footnote{The method for formal reliability assessment is presented
in~\cite{HDC2016JAL} in details.}. For example, the reliability computed from
Option-$(1A)$ in Table~\ref{table: acc_r1} is given as,
$[1 - (1 - R_{\tt thrt\_adj})^2] \times [1 - (1 - R_{brk\_adj})^2] = 0.9504$,
since there are $2$ spatial redundancy provisions for each action in ${\tt
ACC\_R1}$ with respect to the outcomes of ${\tt ACC\_C1}$. Similarly, the
reliability computed from Option-$(2B)$ in Table~\ref{table: acc_r2} is given
as, $[1 - (1 - R_{\tt thrt\_adj} \times R_{brk\_adj})^5] = 0.9983$, since there
are $5$ ways to satisfy ${\tt ACC\_C2}$ from the given action-strategy of
${\tt ACC\_R1}$.

\begin{table*}[htb]
\caption{Possible Options of Action-Events for ${\tt ACC\_R1}$
\label{table: acc_r1}}
\centering
\begin{tabular}{|c||c|c|c|c||c|}
\hline
Possible & \multicolumn{4}{|c||}{Action Events (Cycle-wise)} & Computed\\
\cline{2-5}
Options & Cycle-1 & Cycle-2 & Cycle-3 & Cycle-4 & Reliability\\
\hline \hline
\rowcolor{lightgray}
(1A) & ${\tt act1}$ & ${\tt act2}$ & & & $0.9504$\\
\rowcolor{lightgray}
& ${\tt act1}$ & ${\tt act2}$ & & & \\
\hline
\rowcolor{lightgray}
(1B) & ${\tt act1}$ & & ${\tt act2}$ & & $0.9504$\\
\rowcolor{lightgray}
& ${\tt act1}$ & & ${\tt act2}$ & & \\
\hline
\rowcolor{lightgray}
(1C) & & ${\tt act1}$ & ${\tt act2}$ & & $0.9504$\\
\rowcolor{lightgray}
& & ${\tt act1}$ & ${\tt act2}$ & & \\
\hline
\rowcolor{lightgray}
(1D) & & ${\tt act1}$ & & ${\tt act2}$ & $0.9504$\\
\rowcolor{lightgray}
& & ${\tt act1}$ & & ${\tt act2}$ & \\
\hline
\end{tabular}
\end{table*}

\begin{table*}[htb]
\caption{Possible Options of Action-Events for ${\tt ACC\_R2}$
\label{table: acc_r2}}
\centering
\begin{tabular}{|c||c|c|c|c|c|c||c|}
\hline
Possible & \multicolumn{6}{|c||}{Action Events (Cycle-wise)} & Computed\\
\cline{2-7}
Options & Cycle-1 & Cycle-2 & Cycle-3 & Cycle-4 & Cycle-5 & Cycle-6 &
Reliability\\
\hline \hline
\rowcolor{lightgray}
(2A) & ${\tt act1}$ & ${\tt act1}$ & ${\tt act2}$ & & & & $0.9999$\\
\rowcolor{lightgray}
& & ${\tt act1}$ & ${\tt act1}$ & ${\tt act2}$ & & & \\
\hline
\rowcolor{lightgray}
(2B) & ${\tt act1}$ & ${\tt act1}$ & ${\tt act2}$ & & & & $0.9983$\\
\rowcolor{lightgray}
& & & ${\tt act1}$ & ${\tt act1}$ & ${\tt act2}$ & & \\
\hline
(2C) & ${\tt act1}$ & ${\tt act1}$ & ${\tt act2}$ & ${\tt act1}$ & ${\tt act1}$
& ${\tt act2}$ & $0.9216$\\
\hline
\rowcolor{lightgray}
(2D) & & ${\tt act1}$ & ${\tt act1}$ & ${\tt act2}$ & & & $0.9999$\\
\rowcolor{lightgray}
& & & ${\tt act1}$ & ${\tt act1}$ & ${\tt act2}$ & & \\
\hline
(2E) & & ${\tt act1}$ & ${\tt act1}$ & ${\tt act2}$ & & & $0.9780$\\
& & & & ${\tt act1}$ & ${\tt act1}$ & ${\tt act2}$ & \\
\hline
(2F) & & & ${\tt act1}$ & ${\tt act1}$ & ${\tt act2}$ & & $0.9216$\\
& & & & ${\tt act1}$ & ${\tt act1}$ & ${\tt act2}$ & \\
\hline
\end{tabular}
\end{table*}

The \colorbox{lightgray}{highlighted} rows of Table~\ref{table: acc_r1} and 
Table~\ref{table: acc_r2} indicate the action-strategies that meet the desired 
reliability requirements for both properties of ${\tt ACC}$ subsystem. We call 
all these strategies meeting the reliability targets as the {\em admissible} 
action-strategies.
} \qed
\end{example}
}

\section{Reliability-aware Resource Estimation} \label{sec: rsrc_est}
In this section, we formally present the problem of reliability-aware resource
estimation and illustrate the detailed resource allocation procedure.

\subsection{Formal Statement of the Problem}
The problem of finding the optimal number of computing resources (processors)
considering the simultaneous occurrences of a set of sensed-events is formally
described as:

\noindent {\em Given}:
\begin{enumerate}[(i)]
 \item A set of formal correctness and reliability properties of the system,
 
 \item Assigned reliability targets with respect to each correctness 
	specification,

 \item The set of sensed-events that may occur simultaneously, and

 \item A set of admissible action-strategies corresponding to every sensed-event
	(derived from the corresponding reliability properties of the system).
\end{enumerate}

\noindent {\em Objective}:
To determine the appropriate choice of action-strategy with respect to every
sensed-event so that the overall resource/processor requirement is minimized.

\noindent {\em Assumptions}:
\begin{enumerate}[(i)]
 \item Same actions participating under two different action-strategies
	corresponding to different sensed-events can be executed together in the 
	same processor.

 \item Execution of every action takes unit time (one cycle) in the processor.
 
 \item Within one action-strategy, the participating actions appear as
	disjunction-free.
 
\end{enumerate}

Assumption-(i) prevents this resource optimization problem to have similar
behavior as any conventional two-dimensional strip packing 
problems~\cite{LMM2002EJOR} and hence we cannot directly apply those solutions 
here. However, the solution space is determined by several possible choices of 
the action-strategy combinations attributing to the varied number of processors 
required. The following example illustrated this in details.

{\small
\begin{example} \label{ex: acc_rsrc}
{\em Let us revisit Example~\ref{ex: acc_rel} where we derive the admissible 
action-strategies for the two specifications of ${\tt ACC}$ subsystem in the 
\colorbox{lightgray}{highlighted} rows of Table~\ref{table: acc_r1} and 
Table~\ref{table: acc_r2}.

Suppose, the sensed-events, ${\tt lead\_obs}$ and ${\tt lead\_gap}$, occur 
simultaneously at Cycle-0. Then, Table~\ref{table: acc_rsrc} denotes possible 
execution options for the action-events combining the both the strategies. For 
example, if we try to ascertain Options $(1A)$ with $(2A)$, then we need 
{\em two} processors in Cycle-1 to execute two parallel ${\tt act1}$ events 
among which one ${\tt act1}$ event of Option $(1A)$ is shared/paired with the 
${\tt act1}$ from Option $(2A)$.

\begin{table*}[htb]
\small
\caption{Possible Executions of Action-Events and Required Resources for ${\tt
ACC}$ Subsystem \label{table: acc_rsrc}}
\centering
\begin{tabular}{|c||c|c|c|c|c||c|}
\hline
Strategy & \multicolumn{5}{|c||}{Possible Executions of Action-Events
(Cycle-wise)} & Resource\\
\cline{2-6}
Comb. & Cycle-1 & Cycle-2 & Cycle-3 & Cycle-4 & Cycle-5 & Req.\\
\hline \hline
1A+2A & $\langle {\tt act1}, {\tt act1} \rangle$ & $\langle {\tt act1}, {\tt
act1}, {\tt act2}, {\tt act2} \rangle$ & $\langle {\tt act1}, {\tt act2}
\rangle$ & $\langle {\tt act2} \rangle$ & & $4$\\
\hline
1A+2B & $\langle {\tt act1}, {\tt act1} \rangle$ &  $\langle {\tt act1}, {\tt
act2}, {\tt act2} \rangle$ & $\langle {\tt act1}, {\tt act2} \rangle$ & $\langle
{\tt act1} \rangle$ & $\langle {\tt act2} \rangle$ & $3$\\
\hline
1A+2D & $\langle {\tt act1}, {\tt act1} \rangle$ & $\langle {\tt act1}, {\tt
act2}, {\tt act2} \rangle$ & $\langle {\tt act1}, {\tt act1}\rangle$ & $\langle
{\tt act1}, {\tt act2} \rangle$ & $\langle {\tt act2} \rangle$ & $3$\\
\hline
1B+2A & $\langle {\tt act1}, {\tt act1} \rangle$ & $\langle {\tt act1}, {\tt
act1} \rangle$ & $\langle {\tt act1}, {\tt act2}, {\tt act2} \rangle$ & $\langle
{\tt act2} \rangle$ & & $3$\\
\hline
1B+2B & $\langle {\tt act1}, {\tt act1} \rangle$ & $\langle {\tt act1} \rangle$
& $\langle {\tt act1}, {\tt act2}, {\tt act2} \rangle$ & $\langle {\tt act1}
\rangle$ & $\langle {\tt act2} \rangle$ & $3$\\
\hline
1B+2D & $\langle {\tt act1}, {\tt act1} \rangle$ & $\langle {\tt act1} \rangle$
& $\langle {\tt act1}, {\tt act1}, {\tt act2}, {\tt act2} \rangle$ & $\langle
{\tt act1}, {\tt act2} \rangle$ & $\langle {\tt act2} \rangle$ & $4$\\
\hline
1C+2A & $\langle {\tt act1} \rangle$ & $\langle {\tt act1}, {\tt
act1} \rangle$ & $\langle {\tt act1}, {\tt act2}, {\tt act2} \rangle$ & $\langle
{\tt act2} \rangle$ & & $3$\\
\hline
1C+2B & $\langle {\tt act1} \rangle$ & $\langle {\tt act1}, {\tt act1} \rangle$
& $\langle {\tt act1}, {\tt act2}, {\tt act2} \rangle$ & $\langle {\tt act1}
\rangle$ & $\langle {\tt act2} \rangle$ & $3$\\
\hline
1C+2D & & $\langle {\tt act1}, {\tt act1} \rangle$ & $\langle {\tt act1}, {\tt
act1}, {\tt act2}, {\tt act2} \rangle$ & $\langle {\tt act1}, {\tt act2}
\rangle$ & $\langle {\tt act2} \rangle$ & $4$\\
\hline
1D+2A & $\langle {\tt act1} \rangle$ & $\langle {\tt act1}, {\tt act1} \rangle$
& $\langle {\tt act1}, {\tt act2} \rangle$ & $\langle {\tt act2}, {\tt act2}
\rangle$ & & $2$\\
\hline
1D+2B & $\langle {\tt act1} \rangle$ & $\langle {\tt act1}, {\tt act1} \rangle$
& $\langle {\tt act1}, {\tt act2} \rangle$ & $\langle {\tt act1}, {\tt act2},
{\tt act2} \rangle$ &  $\langle {\tt act2} \rangle$ & $3$\\
\hline
1D+2D & & $\langle {\tt act1}, {\tt act1} \rangle$ & $\langle {\tt act1}, {\tt
act1} \rangle$ & $\langle {\tt act1}, {\tt act2}, {\tt act2} \rangle$ & $\langle
{\tt act2} \rangle$ & $3$\\
\hline
\end{tabular}
\end{table*}

Such action sharing helps us to reduce the number of processor requirements. We 
find that $(1A+2A)$ combination requires $4$ processors to execute the action 
strategies. Table~\ref{table: acc_rsrc} presents all the possible combinations 
and the required resources while executing each of these combinations of action 
strategies. It may be noted that, the minimum number of resources ($2$) is 
required when we perform the actions as per the Options $(1D)$ and $(2A)$ for 
both the properties. Hence, the next challenge is to bind the action-events
with respect to Cycles so that the required resources are minimized.
} \qed
\end{example}
}

\subsection{Resource Estimation Procedure}
The required number of resources can be computed from a set of admissible
action-strategies (corresponding to sensed-events) derived from the reliability
properties of a system, assuming that the corresponding set of sensed-events
occur simultaneously. The minimum count of execution units depends on the 
optimal allocation of the actions to appropriate processor cores in every 
execution time/cycle so as to maximize the sharing of action executions. The 
entire problem can be modeled as a Constraint Satisfaction Problem (CSP) which 
can be solved by an SMT / ILP solver.

\begin{figure}[htb] 
  \begin{center}
  \includegraphics[scale=0.5]{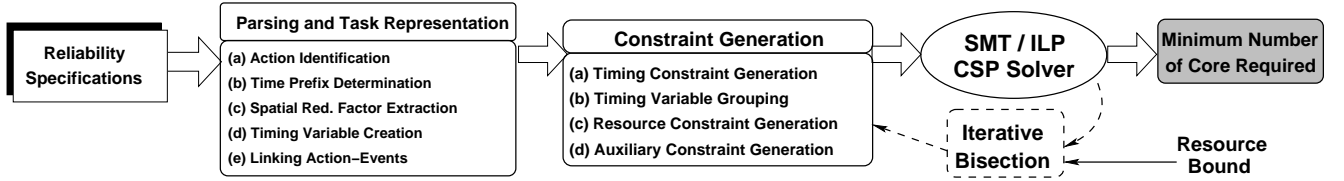}
  \end{center} 
  \caption{Resource Estimation Framework}
  \label{fig: framework}
\end{figure}

Figure~\ref{fig: framework} illustrates the primary steps in this framework. 
The steps that are involved here are primarily categorized in two broad parts, 
namely -- {\em 1. Parsing and Action Representation} and {\em 2. Constraint
Generation}.

\subsubsection*{\bf 1. Parsing and Action Representation}
From the reliability properties and the derived action-strategies, the first
step is to identify relevant information of the action-event occurrences, their
timing and redundancy related information so that the required constraints can
be derived later. We present various stages of parsing and action representation
as follows.
\begin{enumerate}[(a)]
 \item {\em Action Identification:}
	This step identifies the set of action-events from reliability
	specifications. For reliability specifications having redundant actions,
	duplicate action-events are created.

 \item {\em Time Prefix Determination:}
	The time prefix of an action-event, extracted from the reliability
	specification, is typically given as either $\text{\#\#} t_0$
	(fixed-time) or $\text{\#\#} [t_1:t_2]$ (time-range) where $t_0, t_1,
	t_2 \in \mathbb{N}$, the set of all non-negative integers. The lower and
	upper bound of the time prefixes are represented using a doublet,
	$\langle t_0, t_0 \rangle$, for the fixed time prefix and a doublet,
	$\langle t_1, t_2 \rangle$, for representing the variable time prefix.
	The redundancies are treated as follows:
	\begin{itemize}
	 \item In case of spatial redundancy ($m$-times), all replicated actions
		(from $2^{nd}$ to $m^{th}$ occurrences) have $\langle 0, 0
		\rangle$ as the time-prefix.

	\item In case of temporal redundancy ($k$-times), let the time prefix
		for the first action in the first execution be extracted as
		$\langle t_1, t_2 \rangle$. Then, the first action of $i^{th}$
		($i \in [1,k]$) re-executed sequence can start earliest at
		$(t_1+i)$ and latest by $\Delta$\footnote{$\Delta$ is computed
		from the reliability specification considering the given
		reliability target of the corresponding correctness property.
		Intuitively, this upper bound in time comes from the fact that
		delaying the start (beyond $\Delta$) in re-executing the same
		sequence reduces the number of possible satisfaction of the
		correctness property and hence the specified reliability remains
		unmet (Literature~\cite{HDC2016JAL} provides more details).} --
		thereby having $\langle (t_1+i), \Delta \rangle$ as the
		time-prefix. The time prefixes of the other actions are
		extracted from the specification and remain same across all
		their re-executions.
	\end{itemize}

 \item {\em Spatial Redundancy Factor Extraction:}
	Action-events with spatial redundancy of $m$ are represented by
	associating each replication with a spatial redundancy factor from $1$
	to $m$. A default spatial redundancy factor of $1$ is assigned to
	action-events having no spatial redundancy.

 \item {\em Timing Variable Creation:}
 	Timing variable represents the time of occurrence of an action-event. 
	For every action-event identified from the reliability properties, 
	unique timing variables are created for each occurrence (replicated or 
	re-executed) of that action, which will be used to derive the 
	constraints relating to the time of execution of these action-events.

 \item {\em Linking Action-Events:}
	For every action-event appearing in the reliability specification, the
	timing-related constraints will be generated either from the absolute
	time of its occurrence or in relative to the timing of its preceding
	event. Hence, we need to map each timing variable (corresponding to an
	action-event) to the timing variable of its preceding action -- thereby
	aiding to the generation of timing constraints for action-event
	sequences with respect to an action-strategy. The first action-event in 
	the reliability specification is not linked to any other events, since 
	it depends only on the time of occurrence of the sensed-event. Hence, 
	the timing variable corresponding to the first action-event has its 
	linked variable as $\phi$. For redundancies, the action-event links are 
	done as follows:
	\begin{itemize}
	 \item In case of spatial redundancy ($m$-times), all replicated actions
		(from $2^{nd}$ to $m^{th}$ occurrences) are linked with the
		first occurrence of that action.

	 \item In case of temporal redundancy ($k$-times), the first action in
		the $i^{th}$ re-execution ($i \in [1,k]$) is linked with the
		first action in the $(i-1)^{th}$ re-execution.
	\end{itemize}
\end{enumerate}
These stages of parsing and action representation are illustrated in details
through Example~\ref{ex: acc_parse}.

{\small
\begin{example} \label{ex: acc_parse}
{\em Consider the reliability specifications, ${\tt ACC\_R1}$ and ${\tt 
ACC\_R2}$, with their corresponding correctness specifications, ${\tt ACC\_C1}$ 
and ${\tt ACC\_C2}$, as given in Example~\ref{ex: acc_rel}.
\begin{table*}[htb]
\small
\caption{Action-event Information extracted after Parsing and Action 
Representation Stages}
\label{table: info_acc_r1_r2}
\subfloat[Action-event Information Extracted for ${\tt ACC\_R1}$]
{
\begin{tabular}{|c|c|c|c|c|}
\hline
Action & Time Pref. & Sp.-Red. Factor & Time Var. & Link Var.\\
\hline \hline
${\tt act1}$ & $\langle 1, 2 \rangle$ & $1$ & $\tau_{11}$ & $\phi$\\
\hline
${\tt act1}$ & $\langle 0, 0 \rangle$ & $2$ & $\tau_{12}$ & $\tau_{11}$\\
\hline
${\tt act2}$ & $\langle 1, 2 \rangle$ & $1$ & $\tau_{13}$ & $\tau_{12}$\\
\hline
${\tt act2}$ & $\langle 0, 0 \rangle$ & $2$ & $\tau_{14}$ & $\tau_{13}$\\
\hline
\multicolumn{5}{c}{}\\
\multicolumn{5}{c}{$^\dagger \Delta$ is computed from ${\tt ACC\_R2}$ w.r.t. the
reliability targets of ${\tt ACC\_C2}$}
\end{tabular}
}
\subfloat[Action-event Information Extracted for ${\tt ACC\_R2}$]
{
\begin{tabular}{|c|c|c|c|c|}
\hline
Action & Time Pref. & Sp.-Red. Factor & Time Var. & Link Var.\\
\hline \hline
${\tt act1}$ & $\langle 1, 3 \rangle$ & $1$ & $\tau_{21}$ & $\phi$\\
\hline
${\tt act1}$ & $\langle 1, 1 \rangle$ & $1$ & $\tau_{22}$ & $\tau_{21}$\\
\hline
${\tt act2}$ & $\langle 1, 1 \rangle$ & $1$ & $\tau_{23}$ & $\tau_{22}$\\
\hline
${\tt act1}$ & $\langle 1, \Delta^\dagger \rangle$ & $1$ & $\tau_{24}$ &
$\tau_{21}$\\
\hline
${\tt act1}$ & $\langle 1, 1 \rangle$ & $1$ & $\tau_{25}$ & $\tau_{24}$\\
\hline
${\tt act2}$ & $\langle 1, 1 \rangle$ & $1$ & $\tau_{26}$ & $\tau_{25}$\\
\hline
\end{tabular}
}
\end{table*}
Table~\ref{table: info_acc_r1_r2}(a) and Table~\ref{table: info_acc_r1_r2}(b) 
show the information extracted for these properties after {\em parsing and 
action representation} stages, assuming that the sensed-events, ${\tt 
lead\_obs}$ and ${\tt lead\_gap}$, for the properties happen together at time 
$t=0$.
} \qed
\end{example}
}
 
\subsubsection*{\bf 2. Constraint Generation}
Primarily, the set of generated constraints can be of the following two types --
(i) timing-related constraints, and (ii) resource-related constraints. However,
the detailed stages, needed to derive these constraints to be used by the CSP
solver, are given below.
\begin{enumerate}[(a)]
 \item {\em Timing Constraint Generation:}
	Given a reliability specification, ${\tt Spec_i}$, let the timing
	variable corresponding to an action, ${\tt act_j}$, be $\tau_{ix}$. It
	has the time-prefix $\langle t_x^1, t_x^2 \rangle$ and is linked with
	another action, ${\tt act_k}$, having the timing variable as $\tau_{iy}$
	($y<x$). Then, the timing constraints for $\tau_{ix}$ are,
	{\small
	\begin{eqnarray}
	  \tau_{ix} \geq
		\begin{cases}
		  t_x^1, & \text{if}\ \tau_{iy}=\phi \\
		  \tau_{iy} + t_x^1, & \text{otherwise}
		\end{cases}\ \ \text{ and }\ \
	  \tau_{ix} \leq
		\begin{cases}
		  t_x^2, & \text{if}\ \tau_{iy}=\phi \\
		  \tau_{iy} + t_x^2, & \text{otherwise}
		\end{cases}
	\end{eqnarray}
	}
	If $t_x^1=t_x^2$, then the timing constraint is simplified as,
	{\small
	\begin{eqnarray}
	  \tau_{ix} =
		\begin{cases}
		  t_x^1, & \text{if}\ \tau_{iy}=\phi \\
		  \tau_{iy} + t_x^1, & \text{otherwise}
		\end{cases}
	\end{eqnarray}
	}

 \item {\em Timing Variable Grouping:}
 	
 	Timing variable are grouped to enable resource minimization through its
	sharing. Variables corresponding to same action are grouped together 
	and corresponding to every action, at least one group is formed.
	Given two reliability specifications, ${\tt Spec_i}$ and ${\tt Spec_j}$,
	having a common action, ${\tt act_k}$, let the timing variables
	corresponding to ${\tt act_k}$, be $\tau_{ix}$ and
	$\tau_{jy}$, respectively. Then, we can put $\tau_{ix}$ and $\tau_{jy}$
	together forming one group, $\mathcal{G}_l$, i.e., $\tau_{ix}, \tau_{jy}
	\in \mathcal{G}_l$. It may be noted that, all the timing variables
	belonging to one group, say $\mathcal{G}_l$, are associated with the
	same action-event, say ${\tt act_k}$, from different properties. Here,
	${\tt act_k}$ can also be the first occurrence of an action being
	replicated $m$-times (spatial redundancy).

	Moreover, in case of spatial redundancy (say, action ${\tt act_s}$ in 
	${\tt Spec_i}$ is replicated $m$-times), the timing variables, 
	$\tau_{i{s_2}}, \tau_{i{s_3}}, \ldots, \tau_{i{s_m}}$, corresponding to 
	every replicated action ($2^{nd}$ to $m^{th}$ occurrence) forms a 
	singleton group, i.e., $\mathcal{G}_{l_2} = \{ \tau_{i{s_2}} \}, 
	\mathcal{G}_{l_3} = \{ \tau_{i{s_3}} \}, \ldots, \mathcal{G}_{l_m} = \{ 
	\tau_{i{s_m}} \}$. Now, if ${\tt act_s}$ exists in another 
	specification, ${\tt Spec_j}$, with spatial redundancy $n$-times, then 
	the grouping varies, depending on the relative values of $n$ and $m$:
	{\small
	\begin{itemize}
	  \item If $n \leq m$, then $\mathcal{G}_{l_k} = \mathcal{G}_{l_k} \cup 
		\{ \tau_{j{s_k}} \}\ (\forall k \in [1,n])$.

	  \item If $n > m$, then $\mathcal{G}_{l_k} = \mathcal{G}_{l_k} \cup \{
		\tau_{j{s_k}} \}\ (\forall k \in [1,m])$ and singleton groups, 
		$\mathcal{G}_{l_k} = \{ \tau_{j{s_k}} \}\ (\forall k \in 
		[m+1,n])$.
	\end{itemize}
	}

 \item {\em Resource Constraint Generation:}
	Pairing of timing variables into groups helps us to generate resource
	constraints such that, if $\tau_{ik_1}, \tau_{jk_2} \in \mathcal{G}_l$
	then the corresponding actions (say, ${\tt act_{k_1}}$ from ${\tt
	Spec_i}$ and ${\tt act_{k_2}}$ from ${\tt Spec_j}$) are same, i.e. ${\tt
	act_{k_1} = act_{k_2}}$. So, these actions can be executed once
	whenever possible -- resulting in a reduction in the execution resources
	(processors). Let an action-event, executing at cycle-$t$, be
	represented using the timing variable $\tau_{ij}$; then the required
	resource due to the execution of only that action is given as,
	\begin{equation*}
	 {\tt count}^t(\tau_{ij}) = 
		\begin{cases}
		  1, & \text{if}\ \tau_{ij}=t \\
		  0, & \text{otherwise}
		\end{cases}
	\end{equation*}
	The number of resources required at ${\tt Cycle}$-$t$ by the group of
	timing variables representing an action is denoted as,
	\begin{equation*}
	 {\tt count}^t(\mathcal{G}_l) = 
		\begin{cases}
		  1, & \text{if}\ \exists \tau_{ij} \in \mathcal{G}_l \text{,
		  such that } \tau_{ij}=t\\
		  0, & \text{otherwise}
		\end{cases}
	\end{equation*}
	Suppose we are given with $n$ specifications, ${\tt Spec_1}$, ${\tt 
	Spec_2}$, $\ldots$, ${\tt Spec_n}$. We choose all timing variables 
	belonging to each ${\tt Spec_i}$ ($i \in [1,n]$) from the group,
	\[ \mathcal{G}_l = \lbrace  \tau_{1x_1}, \tau_{1y_1}, \ldots, 
	\tau_{2x_2}, \tau_{2y_2}, \ldots, \tau_{nx_n}, \tau_{ny_n}, \ldots 
	\rbrace \]
	and create $n$ sub-groups of $\mathcal{G}_l$ as,
	\begin{equation*}
	  \mathcal{S}^i_{\mathcal{G}_l} = \{ \tau_{ix_i}, \tau_{iy_i}, \ldots 
	  \},\ \forall i \in [1,n]
	\end{equation*}
	Now, the required resource count for each of these sub-groups, 
	$\mathcal{S}^i_{\mathcal{G}_l}$, is denoted as,
	\begin{eqnarray*}
	  {\tt count}^t(\mathcal{S}^i_{\mathcal{G}_l}) &=& \sum\limits_{\forall
	  \tau_{ij} \in \mathcal{S}^i_{\mathcal{G}_l}} {\tt count}^t(\tau_{ij})
	\end{eqnarray*}
	The sub-group count indicates the number of coincident action-events 
	belonging to the same specification (implied by temporal redundancy). 
	Hence, we indicate the total number of resources required at ${\tt  
	Cycle}$-$t$ to execute the actions from these n sub-groups of
	$\mathcal{G}_l$ as,
	\begin{eqnarray*}
	{\tt count}_{\tt SUB}^t(\mathcal{G}_l) = \mathop{\tt MAX}_{1 \leq i 
	\leq n} [ {\tt count}^t(\mathcal{S}^i_{\mathcal{G}_l}) ] 
	\end{eqnarray*}
	
	Let all the timing variables corresponding to the action-event, ${\tt
	act_k}$, appear in $m$ groups, $\mathcal{G}_{l_1}$, $\mathcal{G}_{l_2}$,
	$\ldots$, $\mathcal{G}_{l_m}$. Then, we define,
	\[ {\tt count}_{\tt SUP}^t(\mathcal{G}_{l_j}) =
	\sum\limits_{j=1}^m {\tt count}^t(\mathcal{G}_{l_j}),\ \forall j \in
	[1,m] \]
	A pertinent point to note here is that, for a group, $\mathcal{G}_l$, if
	there is no temporal redundancy involved (in the specification) for the
	representative action of that group, then we find, $[ {\tt count}_{\tt
	SUB}^t(\mathcal{G}_l) - {\tt count}_{\tt SUP}^t(\mathcal{G}_l) ] \leq 0$
	and the required resources at ${\tt Cycle}$-$t$ becomes, ${\tt count}^t
	= \sum_{l} {\tt count}^t(\mathcal{G}_l)$. Hence, the generated resource
	constraint for ${\tt Cycle}$-$t$ is derived as follows:
	\begin{equation}
	  {\tt count}^t = 
		\begin{cases}
		  \sum\limits_{\forall l} {\tt count}^t(\mathcal{G}_l) \text{
		  \bf; if } [ {\tt count}_{\tt SUB}^t(\mathcal{G}_l) - {\tt
		  count}_{\tt SUP}^t(\mathcal{G}_l) ] \leq 0\\
		  \sum\limits_{\forall l} [ {\tt count}^t(\mathcal{G}_l) + \{
		  {\tt count}_{\tt SUB}^t(\mathcal{G}_l) - {\tt count}_{\tt
		  SUP}^t(\mathcal{G}_l) \}] \text{ \bf; otherwise}
		\end{cases}
	\end{equation}

 \item {\em Auxiliary Constraint Addition:}
	Two types of auxiliary (additional) constraints are derived to make
	the constraint generation process complete. These are discussed below.
	\begin{itemize}
	\item Admissibility Constraint: We have noticed in Section~\ref{sec:
	background} that, among all the action-strategies that are generated
	from a reliability specification with respect to the given reliability
	target of its corresponding correctness property, there is a subset of
	action-strategies which are {\em admissible}. Let us assume that the
	last execution of an action-event can happen latest at ${\tt
	Cycle}$-$\mathcal{T}$ such that all action-strategies remain admissible.
	Then, we need to add a constraint with respect to the timing variable,
	$\tau_{ij}$, corresponding to the last executed action as,
	\[ \tau_{ij} \leq \mathcal{T} \]

	\item Resource Limit Constraint: We start with a pessimistic bound on
	the number of resources required and gradually refine that limit. For a
	given choice of maximum number of resources, $\Gamma \in \mathbb{N}$, 
	these constraints are expressed as,
	\[ {\tt count}^t \leq \Gamma,\ \forall t \in [1,\mathcal{T}] \]
	\end{itemize}
\end{enumerate}

All these generated constraints can also be used in a constraint optimization 
tool/solver (like ILP solvers). There, we need to add the additional objective 
function providing the minimization or maximization requirement. Since we are 
minimizing the number of resources (processors) in this case, the objective 
function will be:
\[ {\tt minimize} (\Gamma) \]

{\small
\begin{example} \label{ex: acc_constraints}
{\em Let us revisit Table~\ref{table: info_acc_r1_r2} generated in
Example~\ref{ex: acc_parse} and derive the required set of constraints.

\begin{itemize}
 \item Timing Constraints: The timing constraints with respect to the property,
	${\tt ACC\_R1}$, are derived as:
	\begin{equation*}
	  \tau_{11} \geq 1;\ \ \ \tau_{11} \leq 2;\ \ \
	  \tau_{13} \geq \tau_{12} + 1;\ \ \ \tau_{13} \leq \tau_{12} + 2;\ \ \
	  \tau_{12} = \tau_{11} + 0;\ \ \ \tau_{14} = \tau_{13} + 0
	\end{equation*} 
	Similarly, the timing constraints with respect to the property, ${\tt
	ACC\_R2}$, are derived as:
	\begin{equation*}
	  \tau_{21} \geq 1;\ \ \ \tau_{21} \leq 2;\ \ \ 
	  \tau_{22} = \tau_{21} + 1;\ \ \ \tau_{23} = \tau_{22} + 1;\ \ \ 
	  \tau_{24} \geq \tau_{21} + 1;\ \ \
	  \tau_{24} \leq \tau_{21} + \Delta;\ \ \
	  \tau_{25} = \tau_{24} + 1;\ \ \ \tau_{26} = \tau_{25} + 1
	\end{equation*}
	Now, $\Delta = 2$ is extracted from the logical satisfaction of ${\tt
	ACC\_C2}$ using ${\tt ACC\_R2}$ subject to specific reliability targets.

 \item Timing Variable Groups: The timing variables for each action-events is
	grouped as follows:
	\begin{equation*}
	  \mathcal{G}_1 = \{ \tau_{11}, \tau_{21}, \tau_{22}, \tau_{24},
	  \tau_{25} \};\ \ \ \mathcal{G}_2 = \{ \tau_{12} \};\ \ \ 
	  \mathcal{G}_3 = \{ \tau_{13}, \tau_{23}, \tau_{26} \};\ \ \ 
	  \mathcal{G}_4 = \{ \tau_{14} \}
	\end{equation*}
	Here, all the timing variables present in the groups $\mathcal{G}_1$ and
	$\mathcal{G}_2$ correspond to ${\tt act1}$ action-event and that are
	present in the groups $\mathcal{G}_3$ and $\mathcal{G}_4$ correspond to
	${\tt act2}$ action-event.

 \item Resource Constraints: The resource constraints are derived using
	Table~\ref{table: rsrc_cons} for each ${\tt Cycle}$-$t$. Finally, we
	produce the following constraints ($\forall t \in [1,5]$):
	\begin{equation*}
	  {\tt count}^t = 
		\begin{cases}
		  \sum\limits_{l=1}^4 {\tt count}^t(\mathcal{G}_l) \text{
		  \bf; if } [ {\tt count}_{\tt SUB}^t(\mathcal{G}_l) - {\tt
		  count}_{\tt SUP}^t(\mathcal{G}_l) ] \leq 0\\
		  \sum\limits_{l=1}^4 [ {\tt count}^t(\mathcal{G}_l) + \{
		  {\tt count}_{\tt SUB}^t(\mathcal{G}_l) - {\tt count}_{\tt
		  SUP}^t(\mathcal{G}_l) \}] \text{ \bf; otherwise}
		\end{cases}
	\end{equation*}

\begin{table}[htb]
\footnotesize
\caption{Resource Constraint Generation (for ${\tt Cycle}$-$t$)}
\centering
\label{table: rsrc_cons}
\begin{tabular}{|c|c|c|c|c|}
\hline
Groups & ${\tt count}^t(\mathcal{G}_l)$ & ${\tt
count}^t(\mathcal{S}^1_{\mathcal{G}_l})$ & ${\tt
count}^t(\mathcal{S}^2_{\mathcal{G}_l})$ & ${\tt count}_{\tt
SUP}^t(\mathcal{G}_l)$\\
\hline \hline
$\mathcal{G}_1$ & \begin{tabular}[c]{@{}c@{}}$(\tau_{11}==t) \lor (\tau_{21}==t) 
\lor$\\ $(\tau_{22}==t) \lor (\tau_{24}==t) \lor (\tau_{25}==t)$\end{tabular} & 
$(\tau_{11}==t)$ & \begin{tabular}[c]{@{}c@{}}$
(\tau_{21}==t) + (\tau_{22}==t) +$\\ $(\tau_{24}==t) + 
(\tau_{25}==t)$\end{tabular} & \begin{tabular}[c]{@{}c@{}}${\tt
count}^t(\mathcal{G}_1) + {\tt count}^t(\mathcal{G}_2)$\end{tabular}\\
\hline
$\mathcal{G}_2$ & $(\tau_{12}==t)$ & $(\tau_{12}==t)$ & $0$ &
\begin{tabular}[c]{@{}c@{}}${\tt count}^t(\mathcal{G}_1) + {\tt
count}^t(\mathcal{G}_2)$\end{tabular}\\
\hline
$\mathcal{G}_3$ & \begin{tabular}[c]{@{}c@{}}$(\tau_{13}==t) \lor (\tau_{23}==t) 
\lor (\tau_{26}==t)$\end{tabular} & $(\tau_{13}==t)$ & 
\begin{tabular}[c]{@{}c@{}}$(\tau_{23}==t) + (\tau_{26}==t)$\end{tabular}
& \begin{tabular}[c]{@{}c@{}}${\tt count}^t(\mathcal{G}_3) + {\tt
count}^t(\mathcal{G}_4)$\end{tabular}\\
\hline
$\mathcal{G}_4$ & $(\tau_{14}==t)$ & $(\tau_{14}==t)$ & $0$ &
\begin{tabular}[c]{@{}c@{}}${\tt count}^t(\mathcal{G}_3) + {\tt
count}^t(\mathcal{G}_4)$\end{tabular}\\
\hline
\end{tabular}
\end{table}

 \item Auxiliary Constraints: The admissibility constraint, here, restricts the
	last action-event, ${\tt act2}$, to happen latest by $5^{th}$-cycle,
	i.e. $\mathcal{T} = 5$ (Refer to Table~\ref{table: info_acc_r1_r2} of 
	Example~\ref{ex: acc_parse} for admissible action-strategies). Hence, we 
	add, $\tau_{26} \leq 5$.
	
	Suppose, we set the resource limits as $\Gamma = 2$, then the added
	constraints are as follows:
	\begin{equation*}
	 {\tt count}^1 \leq 2,\ \ \ {\tt count}^2 \leq 2,\ \ \ {\tt count}^3 
	 \leq 2,\ \ \ {\tt count}^4 \leq 2,\ \ \ {\tt count}^5 \leq 2
	\end{equation*}

\end{itemize}

If we fed all the above constraints together in a SMT/ILP solver, then we find
the following satisfiable valuation of the variables:
\begin{eqnarray*}
\tau_{11}=2, & \tau_{12}=2, & \tau_{13}=4,\ \ \ \tau_{14}=4,\\
\tau_{21}=1, & \tau_{22}=2, & \tau_{23}=3,\ \ \ \tau_{24}=2,\ \ \ \tau_{25}=3,\ 
\ \ \tau_{26}=4.
\end{eqnarray*}
It indicates that, both ${\tt act1}$ and ${\tt act2}$ of ${\tt ACC\_R1}$ is
replicated twice in ${\tt Cycle}$-$2$ and ${\tt Cycle}$-$4$, respectively. Two
consecutive ${\tt act1}$ followed by ${\tt act2}$ of ${\tt ACC\_R2}$ is
executed for the first time in ${\tt Cycle}$-$1$, ${\tt Cycle}$-$2$ and ${\tt
Cycle}$-$3$, respectively. Again, re-execution of the same sequence happens 
when two consecutive ${\tt act1}$ followed by ${\tt act2}$ is executed for
the second time in ${\tt Cycle}$-$2$, ${\tt Cycle}$-$3$ and ${\tt Cycle}$-$4$,
respectively. This result is also evident from Example~\ref{ex: acc_parse},
where the choice of Options $(1D)+(2A)$ produces this outcome as shown in
Table~\ref{table: acc_rsrc}.
} \qed
\end{example}
}

The resource constraints and the timing constraints are fed together to a SMT 
solver to check the satisfiability. If these constraints can be satisfied,
then we conclude that the given resource limit, $\Gamma$, is sufficient for the
admissible action strategies to be scheduled. However, an unsatisfiable outcome
denotes that the resource limit needs further increment. To solve this using an
ILP solver, we also add the objective constraint (for minimization).

Now, to derive the optimal (minimum) number of required resources, we start
from a pessimistic limit and proceed on iteratively bisecting the limit (in a
similar manner as done in binary search technique) until we converge into
finding the minimum value of the resource limit. To illustrate the procedure,
let us assume that we start with a given $\Gamma = 4$ in our example and derive
the satisfiable valuations. Then, we bisect the limit and make $\Gamma = \lfloor
\frac{(1+4)}{2} \rfloor = 2$ and still we can derive satisfiable valuations as
illustrated in Example~\ref{ex: acc_constraints}. Next, when we further bisect
and make $\Gamma = \lfloor \frac{(1+2)}{2} \rfloor = 1$, then the constraints
become unsatisfiable. Hence, we conclude that the minimum number of resources
required to execute the admissible strategies is $2$.

\section{Experimental Results} \label{sec: exp_results}
\begin{figure*}[htb]
\begin{center}
\subfloat[Z3 SMT Solver Response]
 {\includegraphics[scale=0.4]{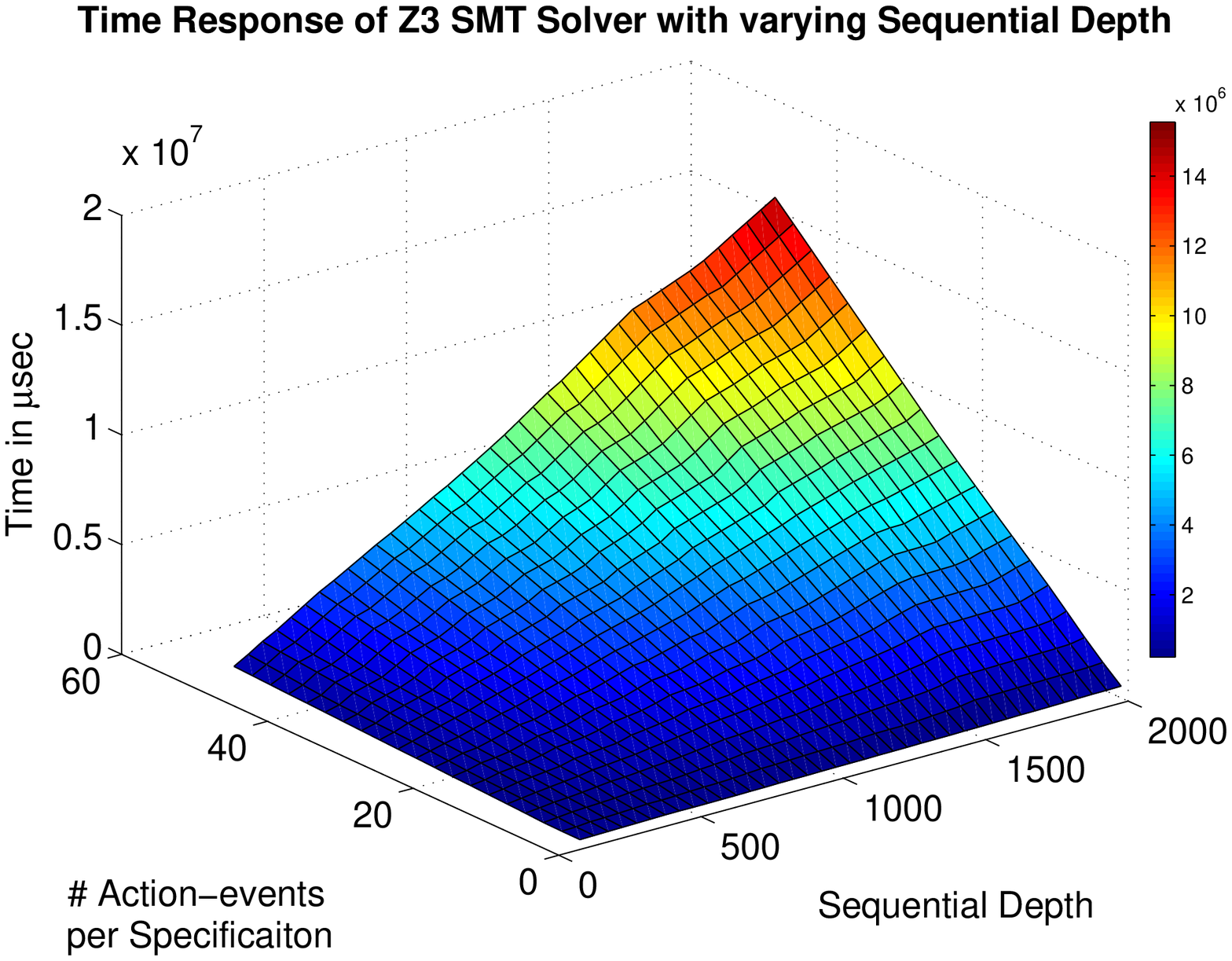}}
\quad
\subfloat[CPLEX ILP Solver Response]
 {\includegraphics[scale=0.4]{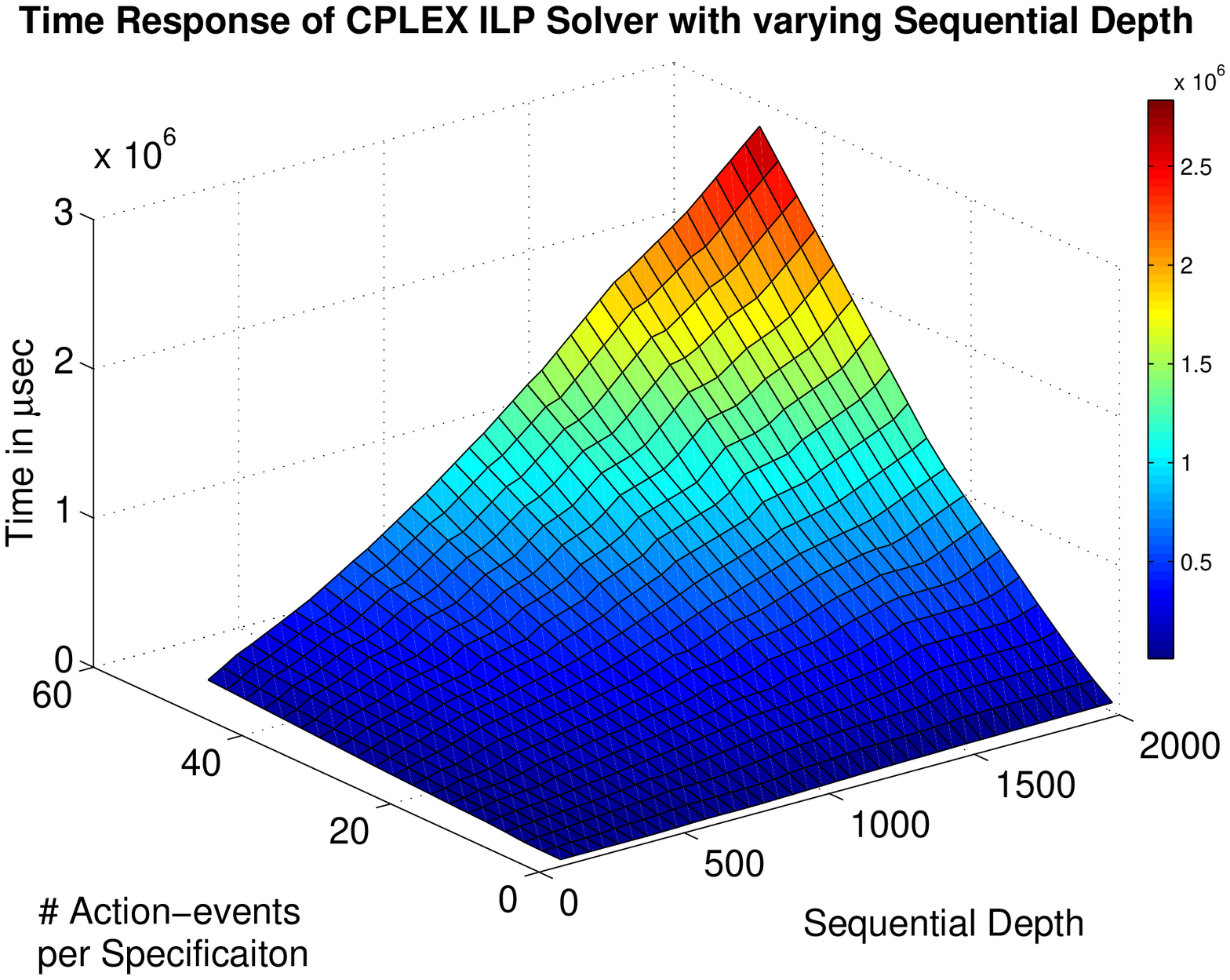}}
\end{center}
\caption{3D Graphs for Scalability Experiments over Sequential Depth of 
Specifications}
\label{fig: 3D_graphs1}
\end{figure*}

\begin{figure*}[htb]
\begin{center}
\subfloat[Z3 SMT Solver Response]
 {\includegraphics[scale=0.23]{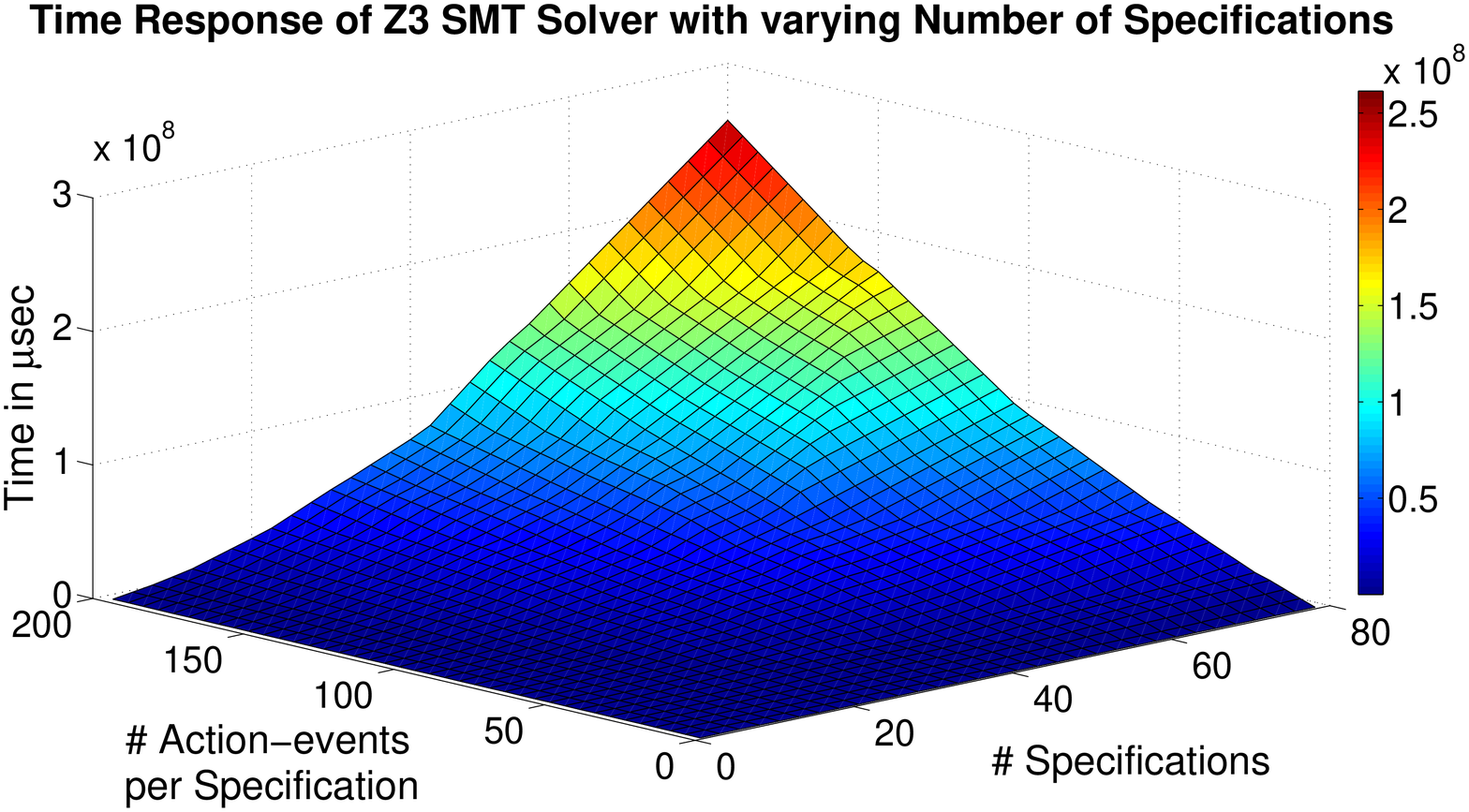}}
\quad
\subfloat[CPLEX ILP Solver Response]
 {\includegraphics[scale=0.23]{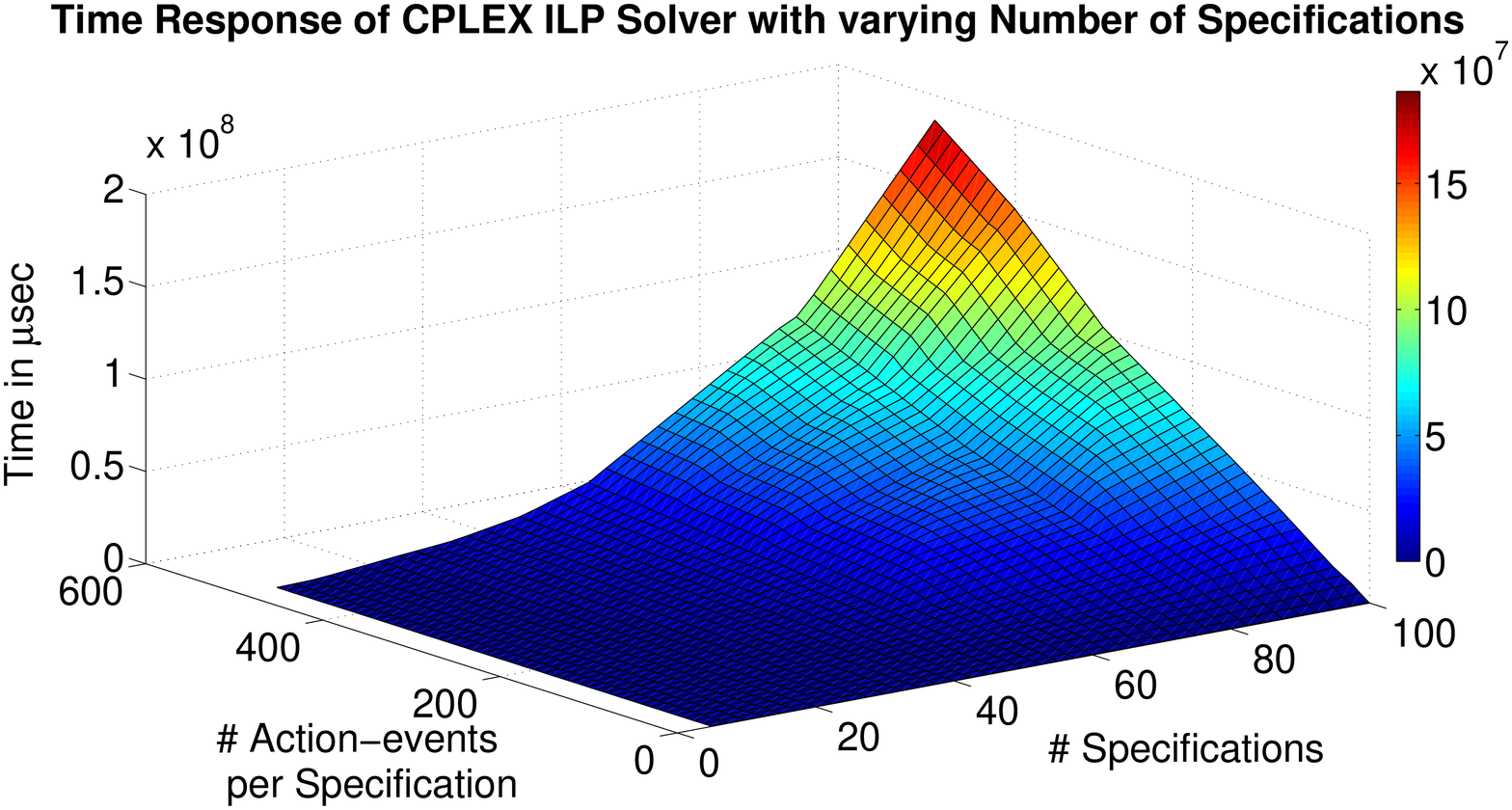}}
\end{center}
\caption{3D Graphs for Scalability Experiments over Number of Specifications}
\label{fig: 3D_graphs2}
\end{figure*}

In this section, the experimental results performed to check the 
{\em scalability} of this approach on an $8$-Core Intel Xeon Ivy bridge 
$E5-2650v2$ series processor ($20M$ Cache, $2.60GHz$) with $4 \times 8GB$ RAM 
are presented. Figure~\ref{fig: 3D_graphs1} and ~\ref{fig: 3D_graphs2} shows 
the 3D plots on the scalability experiments performed. We have used Z3 SMT 
solver~\cite{DB2008TACAS} developed by Microsoft Corporation and CPLEX 
Optimizer (ILP solver)~\cite{CPLEX} developed by IBM. The time response of the 
proposed approach subject to varying sequential depth\footnote{The sequential 
depth of a property is the maximum number of cycles throughout which the 
property may span.} and number of action-events were analyzed. The graphs depict 
exponential increase in execution time in determining the optimal resources as 
the sequential depth of the specifications increase. The time response with 
varying number of properties and varying number of action-events per property 
are also studied. This response also has exponential behavior as indicated in
Figures~\ref{fig: 3D_graphs1}-\ref{fig: 3D_graphs2}.

In Figure~\ref{fig: 3D_graphs1}, the responses of Z3 and CPLEX solvers are 
shown with respect to the variations in sequential depth of the properties. The 
Z-axis in all the graphs represents the time required to find the minimum 
number of resources in Micro-seconds. The X-axis indicates the sequential depth 
(varied upto $2000$ time units) and the Y-axis indicates the number of 
action-events in  the specification (varied upto $50$). For smaller sequential 
depth, the variation in execution time with variation in the number of 
action-events is rather small. However, as the sequential depth increases, the 
execution time varies rapidly as the number of action-events increases. The 
increases in sequential depth also results in the increase in number of 
resource constraints. Moreover, the increase in number of action-events results 
in increase in the length and complexity of resource constraints as well. But 
for same number of action events, the number of timing constraints remains the 
same. The increase in number of resource constraints has higher impact in the 
increase of execution time than the increase in the length of resource 
constraints.

In Figure~\ref{fig: 3D_graphs2}, the response of Z3 and CPLEX solvers are 
shown with respect to the variations in the number of specifications. The 
Z-axis in all the graphs represents the time required to find the minimum 
number of resources in Micro-seconds. The X-axis indicates the number of 
specifications (varied from $0$ to $200$/$500$) and the Y-axis indicates the 
number of action-events per specification (varied upto $80$/$100$). For smaller 
number of specifications, the variation in the execution time with the variation 
in number of action-events is rather small. However, as the number of 
specifications increases, the execution time varies rapidly. Here, the increase 
along the X-axis and Y-axis results in increase in the number of timing 
constraints. This increase in number of timing constraints can result in 
increase of the number of action-events and the number of specifications. There 
will be variation in the number of resource constraints, implied by both 
increase in the number of action-events as well as in the number of properties, 
since time of execution can vary in both these cases. Both X-axis and Y-axis 
has similar effects in increasing the length and number of resource 
constraints. Thus, the behavior is truly exponential in this case. The increase 
in timing constraints have lesser effect on the time of execution than the 
increase in the length/complexity of the resource constraints.

\subsubsection*{Comparison of Results obtained by Constraint Solvers (Z3 and
CPLEX)}
It may be noted that, when modeling our resource estimation problem as a
Constraint Satisfaction Problem (CSP) or a Constraint Optimization Problem
(COP), there are two types of constraints to be considered. The first type is 
the timing constraints that define definite timing relationship between the 
action-events constituting the feasible action-event strategies meeting specific 
reliability targets. The second type is the resource constraint, this restraint 
is to be kept on all viable time value or the maximum depth of time over which 
the action strategy can exist assuming the sensed-event corresponding to the 
reliability specifications happens at time zero.

Both the solvers, Z3 (CSP solver)~\cite{DB2008TACAS} and CPLEX (ILP/COP 
solver)~\cite{CPLEX} will share the same timing and resource constraints. 
However, ILP requires an objective function to be provided in addition to the 
constraints presented to Z3\footnote{In general, the objective function is either
a cost function or energy function which is to be minimized, or a reward function
or utility function, which is to be maximized.}. We can check the satisfiability of 
the given set of reliability specifications with respect to any specific number 
of cores using Z3~\cite{BAR2009SAT}. In CPLEX, the objective function is defined
as the sum of the number of time cycles where the resource constraints are
met and maximization of the same can result in the optimal satisfiable solution
to the problem~\cite{MT2012EUROSTAT}. Hence, we are finding an assignment
that maximizes the number of satisfied resource constraints in ILP to model the
problem.
 
Comparing the performance of the Z3 solver with the CPLEX solver, the CPLEX is 
significantly faster in both scenarios. The primary reason behind this behavior is
the fact that CPLEX ILP solver is a COP solver~\cite{LG2011AIDAC}. The search
space of the ILP solver is limited to the optimal hyper-plane which maximizes
the objective function and hence the ILP solver will be faster in reaching the
minimum number of resources. The calculation of the optimal plane offsets the
difference slightly, still there is an appreciable variation in execution time
between two solvers.

\section{Case-Study: Navigation, Guidance and Control (NGC) System for
Satellite Launch Vehicles} \label{sec: case-study}
The satellite launch vehicle navigation, guidance and control (NGC) system is
responsible for directing the propulsive forces and stabilizing the vehicle to
achieve the orbit with the specified accuracy. The dispersions in propulsion
parameters compared to expected values, variations of the estimated
aerodynamic characteristics during the atmospheric flight phase, the fluctuating
winds both at ground and upper atmosphere and a variety of internal and
external disturbances during the flight can increase the loads on the vehicle
beyond the permissible limit and also cause deviations in the vehicle trajectory
from its intended path~\cite{SS2015BOOK}. Under such environment, NGC
system has to define the  optimum trajectory in real time to reach the specified
target and steer the vehicle along the desired path and inject the spacecraft
into the mission targeted orbit within the specified dispersions while ensuring
the vehicle loads remain within limits. NGC system is among the most crucial
and challenging field in launch vehicle design. The objectives of a satellite
launch vehicle NGC system include:
\begin{enumerate}
 \item Stabilization of the vehicle during its flight against various 
disturbance forces and moments due to aerodynamics, thrust misalignment, 
separation, liquid slosh, etc.
 \item Tracking the desired attitude of the vehicle so as to follow the desired 
trajectory with the specified accuracy to meet the final mission objectives.
 \item Receiving navigation data from sensors (navigation sensors) and desired 
attitude from guidance system and generating appropriate control commands. 
\end{enumerate}

Using the control commands generated by the autopilot system, the actuation 
system generates the necessary forces and moments to control and stabilize the 
vehicle as per the requirements. The inputs to the launch vehicle autopilot 
system are -- navigation sensor data and health status, guidance data and health 
status, sequencing command execution status, feedback from vehicle autopilot 
filters, and feedback from Autopilot on the status of computation and 
validation.

\subsection{Correctness Specifications}
The list of correctness specifications\footnote{\em It is to be noted that a 
single time-unit delay is considered to be of $20$ {\tt ms} in all the above 
mentioned correctness specifications.} for NGC system are expressed formally as 
follows:
{\small
\begin{enumerate}[(1)]
 \item ${\tt NGCS\_C1}$:
{\em If the system detects permanent navigation sensor failure in all channels, 
then the system will discard navigation sensor data within $20$-$40$ {\tt ms},
followed by switching to open loop guidance within $20$-$60$ {\tt ms} and then 
initialize the salvage profile for control within $20$-$60$ {\tt ms}}.
\begin{eqnarray*}
{\tt NGCS\_C1:\ sensor\_failure\_full} & \rightarrow & {\tt \text{\#\#}[1:2]\ 
discard\_sensor\_data}\\
& & {\tt \text{\#\#}[1:3]\ switch\_to\_open\_loop}\\
& & {\tt \text{\#\#}[1:3]\ salvage\_profile\_init}
\end{eqnarray*}

 \item ${\tt NGCS\_C2}$:
{\em If the system detects permanent failure in one of the navigation sensor 
channels, then the system discards failed channel within $20$-$60$ {\tt ms},
followed by recalculation of navigation parameters from redundant channels 
within $20$-$40$ {\tt ms}}.
\begin{eqnarray*}
{\tt NGCS\_C2:\ sensor\_failure\_partial} & \rightarrow & {\tt \text{\#\#}[1:3]\ 
discard\_error\_ch}\\
& & {\tt \text{\#\#}[1:2]\ recal\_navig\_param}
\end{eqnarray*}

 \item ${\tt NGCS\_C3}$:
{\em If the system detects off-nominal navigation parameters from navigation 
software and the navigation sensors being healthy, then the system discards 
present parameters within $20$-$40$ {\tt ms}, followed by recalculation of 
navigation parameters within $20$-$80$ {\tt ms}}.
\begin{eqnarray*}
{\tt NGCS\_C3:\ off\_nominal\_param} & \rightarrow & {\tt \text{\#\#}[1:2]\ 
discard\_pre\_navig\_param}\\
& & {\tt \text{\#\#}[1:4]\ recal\_navig\_param}
\end{eqnarray*}

 \item ${\tt NGCS\_C4}$:
{\em If the system detects guidance algorithm failure, then the system will 
discard present guidance computations within $20$-$60$ {\tt ms}, followed by 
switching to open loop guidance within $20$-$60$ {\tt ms}, followed by salvage 
profile initialization for control within $20$-$40$ {\tt ms}}.
\begin{eqnarray*}
{\tt NGCS\_C4:\ guid\_alg\_failure} & \rightarrow & {\tt \text{\#\#}[1:3]\ 
discard\_present\_comp}\\
& & {\tt \text{\#\#}[1:3]\ switch\_to\_open\_loop}\\
& & {\tt \text{\#\#}[1:2]\ salvage\_profile\_init}
\end{eqnarray*}

 \item ${\tt NGCS\_C5}$:
{\em If the system detects off-nominal values for one of the guidance parameters 
with the guidance algorithm being healthy, then the system discards and flushes 
out the present computation within $20$ {\tt ms}, followed by recalculation 
within $20$-$40$ {\tt ms}}.
\begin{eqnarray*}
{\tt NGCS\_C5:\ guid\_comp\_error} & \rightarrow & {\tt \text{\#\#}1\ 
discard\_present\_comp}\\
& & {\tt \text{\#\#}[1:2]\ recal\_guid\_param}
\end{eqnarray*}

 \item ${\tt NGCS\_C6}$:
{\em If the system detects error in the execution of vehicle autopilot commands, 
then the system switches to open loop guidance within $20$-$60$ {\tt ms}, 
followed by salvage profile initialization for control within $20$-$80$ {\tt 
ms}}.
\begin{eqnarray*}
{\tt NGCS\_C6:\ comm\_exec\_error} & \rightarrow & {\tt \text{\#\#}[1:3]\ 
switch\_to\_open\_loop}\\
& & {\tt \text{\#\#}[1:4]\ salvage\_profile\_init}
\end{eqnarray*}

 \item ${\tt NGCS\_C7}$:
{\em If the system detects any failure in switching Vehicle Autopilot digital 
filters, then the system switches to open loop guidance within $20$-$100$ {\tt 
ms}, followed by salvage profile initialization for control within $20$-$60$ 
{\tt ms}}.
\begin{eqnarray*}
{\tt NGCS\_C7:\ filter\_switch\_fail} & \rightarrow & {\tt \text{\#\#}[1:5]\ 
switch\_to\_open\_loop}\\
& & {\tt \text{\#\#}[1:3]\ salvage\_profile\_init}
\end{eqnarray*}

\item ${\tt NGCS\_C8}$:
{\em If the system detects non-execution of sequencing(time-based) commands for 
the first time, then the system re-schedule time-based commands within $20$-$60$ 
{\tt ms}, followed by rechecking the execution status within $20$-$80$ {\tt 
ms}}.
\begin{eqnarray*}
{\tt NGCS\_C8:\ seq\_comm\_non\_exe} & \rightarrow & {\tt \text{\#\#}[1:3]\ 
reschedule\_seq\_comm}\\
& & {\tt \text{\#\#}[1:4]\ recheck\_seq\_comm\_exec\_status}
\end{eqnarray*}

 \item ${\tt NGCS\_C9}$:
{\em If the system detects non-execution of sequencing(time-based) commands more 
than once, then the system will switch to open loop guidance within $20$-$140$ 
{\tt ms}}.
\begin{eqnarray*}
{\tt NGCS\_C9:\ seq\_comm\_non\_exe} & \rightarrow & {\tt \text{\#\#}[1:7]\ 
switch\_to\_open\_loop}
\end{eqnarray*}

\item ${\tt NGCS\_C10}$:
{\em If the system detects error in on board arithmetic computation unit, then 
the system discards all computations within $20$-$40$ {\tt ms}, followed by 
switching to open loop guidance within $20$-$80${\tt ms}, followed by salvage 
profile initialization for control within $20$-$60${\tt ms}}.
\begin{eqnarray*}
{\tt NGCS\_C10:\ onboard\_comp\_error} & \rightarrow & {\tt \text{\#\#}[1:2]\ 
discard\_present\_comp}\\
& & {\tt \text{\#\#}[1:4]\ switch\_to\_open\_loop}\\
& & {\tt \text{\#\#}[1:3]\ salvage\_profile\_init}
\end{eqnarray*}

 \item ${\tt NGCS\_C11}$:
{\em If the system detects off-normal control parameters with the control 
algorithm being healthy then the system will recalculate guidance parameters 
within $20$-$60$ {\tt ms}, followed by the recalculation of control 
parameters within $20-60${\tt ms}}.
\begin{eqnarray*}
{\tt NGCS\_C11:\ abnormal\_control\_param} & \rightarrow & {\tt \text{\#\#}[1:3]\ 
recal\_guid\_param}\\
& & {\tt \text{\#\#}[1:3]\ recal\_control\_param}
\end{eqnarray*}

 \item ${\tt NGCS\_C12}$:
{\em If the system detects any communication error between navigation, guidance, 
control or sequencing then the system will switch to open loop guidance within 
$20$-$120$ {\tt ms}, followed by salvage profile initialization for control 
within $20$-$60$ {\tt ms}}.
\begin{eqnarray*}
{\tt NGCS\_C12:\ communication\_error} & \rightarrow & {\tt \text{\#\#}[1:6]\ 
switch\_to\_open\_loop}\\
& & {\tt \text{\#\#}[1:3]\ salvage\_profile\_init}
\end{eqnarray*}

 \item ${\tt NGCS\_C13}$:
{\em If the system detects temporary failure in any of the navigation sensor 
channels then the system will discard sensor data within $20$-$40$ {\tt ms},
followed by the recalculation of navigation parameters using the redundant 
channels within $20-80${\tt ms}}.
\begin{eqnarray*}
{\tt NGCS\_C13:\ sensor\_ch\_temp\_failure} & \rightarrow & {\tt \text{\#\#}[1:2]\ 
discard\_sensor\_data}\\
& & {\tt \text{\#\#}[1:4]\ recal\_navig\_param}
\end{eqnarray*}

 \item ${\tt NGCS\_C14}$:
{\em If the system has any temporary failure reported in any of the navigation 
sensor channels then the system will recheck the particular channel values for 
re-induction within $20$-$160$ {\tt ms}}.
\begin{eqnarray*}
{\tt NGCS\_C14:\ sensor\_ch\_temp\_failure} & \rightarrow & {\tt \text{\#\#}[1:8]\
recheck\_error\_ch}
\end{eqnarray*}

 \item ${\tt NGCS\_C15}$:
{\em If the system detects any timing failure due to shift in clock frequency 
then the system will discards all computations within $20$-$40$ {\tt ms},
followed by switching to open loop guidance within $20$-$60$ {\tt ms}, followed 
by salvage profile initialization for control within $20$-$60$ {\tt ms}}.
\begin{eqnarray*}
{\tt NGCS\_C15:\ clock\_freq\_shift} & \rightarrow & {\tt \text{\#\#}[1:2]\
discard\_present\_comp}\\
& & {\tt \text{\#\#}[1:3]\ switch\_to\_open\_loop}\\
& & {\tt \text{\#\#}[1:3]\ salvage\_profile\_init}
\end{eqnarray*}
\end{enumerate}
}

\subsection{Reliability Specifications}
Table~\ref{table: info_ngcs} outlines various activities of NGC system and the 
corresponding outcome-events responsible for that. Also, the respective 
action-events for each outcome event and the reliability values for the 
outcomes, i.e. $R_{\tt outcome\text{-}event} = {\tt Prob(outcome\text{-}event\ 
|\ action\text{-}event)}$, are specified.
\begin{table*}[htb]
\tiny
\caption{Action-events Corresponding to the Outcome-Events of NGC system and 
their Reliability Values}
\label{table: info_ngcs}
\centering
\begin{tabular}{|l||c|c|c|}
\hline
Event Description & Outcome-Event & Action-Event & Reliability Values\\
\hline \hline
switching to open loop guidance & ${\tt switch\_to\_open\_loop}$ & ${\tt 
act1}$ & $R_{\tt switch\_to\_open\_loop} = 0.985$\\
\hline
initialization of salvage profile for control & ${\tt salvage\_profile\_init}$ & 
${\tt act2}$ & $R_{\tt salvage\_timing\_init} = 0.984$\\
\hline
discarding error channel & ${\tt discard\_error\_ch}$ & ${\tt act3}$ & $R_{\tt 
discard\_error\_ch}= 0.985$\\
\hline
recalculation of navigation parameters & ${\tt recal\_navig\_param}$ & ${\tt 
act4}$ & $R_{\tt recal\_navig\_param}= 0.983$\\
\hline
discarding previous navigation parameters & ${\tt discard\_pre\_navig\_param}$ 
& ${\tt act5}$ & $R_{\tt discard\_pre\_navig\_param}= 0.986$\\
\hline
discarding present computations & ${\tt  discard\_present\_comp}$ & ${\tt 
act6}$ & $R_{\tt  discard\_present\_comp}= 0.996$\\
\hline
rescheduling sequencing commands & ${\tt reschedule\_seq\_comm}$ & ${\tt 
act7}$ & $R_{\tt reschedule\_seq\_comm}= 0.982$\\
\hline
rechecking sequencing command execution status & ${\tt 
recheck\_seq\_comm\_exec\_status}$ & ${\tt act8}$ & $R_{\tt 
recheck\_seq\_comm\_exec\_status}= 0.986$\\
\hline
recalculating control parameters & ${\tt recal\_control\_param}$ & ${\tt 
act9}$ & $R_{\tt recal\_control\_param}= 0.982$\\
\hline
discarding sensor data & ${\tt discard\_sensor\_data}$ & ${\tt act10}$ & $R_{\tt 
discard\_sensor\_data}= 0.996$\\
\hline
rechecking error channel & ${\tt recheck\_error\_ch}$ & ${\tt act11}$ & $R_{\tt 
recheck\_error\_ch}= 0.982$\\
\hline
recalculating guidance parameters & ${\tt recal\_guid\_param}$ & ${\tt 
act12}$ & $ R_{\tt recal\_guid\_param}= 0.996$\\
\hline
\end{tabular}
\end{table*}

In addition to this, let us assume that the desired reliability of all the 
given correctness requirements be $0.992$. Since the outcomes are unreliable, 
the NGC system must issue the action-events with appropriate redundancy in 
order to meet the desired reliability target, as expressed formally by the 
following reliability specifications\footnote{\em It is to be noted that a 
single time-unit delay is considered to be of $20$ {\tt ms} in all the above 
mentioned reliability specifications.}.
{\small
\begin{enumerate}[(1)]
 \item ${\tt NGCS\_R1}$:
{\em As soon as the navigation sensor permanent failure is detected in all the 
channels, the navigation sensor data is discarded within $20$-$40$ {\tt 
ms} followed by scheduling the switching to open loop guidance action-event in 
two processors parallelly within $20$-$60$ {\tt ms}, followed by the salvage 
profile initialization action which is also scheduled in two processors 
parallelly within next $20$-$60$ {\tt ms}}.
\[
{\tt NGCS\_R1:\ sensor\_failure\_full \rightarrow \text{\#\#}[1:2]\ act10\ 
\text{\#\#}[1:3]\ act1[\sim 2]\ \text{\#\#}[1:3]\ act2[\sim 2]}
\]

 \item ${\tt NGCS\_R2}$:
{\em As soon as the system detects permanent failure in one of the navigation 
sensor channels, system does the action corresponding to discarding failed 
channel within $20$-$40$ {\tt ms} followed by initiating the action to 
recalculate navigation parameters from redundant channels within $20$-$60$ {\tt 
ms}. The latter action is repeated consecutively twice}.
\[
{\tt NGCS\_R2:\ sensor\_failure\_partial \rightarrow \text{\#\#}[1:2]\ act3\ 
\text{\#\#}[1:3]\ act4[\ast 2]} \]

 \item ${\tt NGCS\_R3}$:
{\em As soon as the system detects off-nominal navigation parameters from 
navigation software and the navigation sensors being healthy, system initiates 
action to discard present parameters within $20$-$40$ {\tt ms} followed by 
recalculation action to calculate navigation parameters within $20$-$40$ {\tt 
ms}. The second action is repeated consecutively twice}.
\[
{\tt NGCS\_R3:\ of\_normal\_param \rightarrow \text{\#\#}[1:2]\ act5 
\text{\#\#}[1:2]\  act4[\ast 2]}
\]

 \item ${\tt NGCS\_R4}$:
{\em whenever the system detects guidance algorithm failure, system will perform 
action corresponding to discard present guidance computations within $20$-$40$ 
{\tt ms} followed by switching to open loop guidance action within $20$-$40$ 
{\tt ms} followed by salvage profile initialization for control within $20$-$40$ 
{\tt ms}. The latter two actions are repeated non-consecutively twice}. 
\[
{\tt NGCS\_R4:\ guid\_alg\_failure \rightarrow \text{\#\#}[1:2]\ act6\ 
\text{\#\#}[1:2]\ (act1\ \text{\#\#}[1:2]\ act2)[= 2]}
\]

 \item ${\tt NGCS\_R5}$:
{\em whenever the system detects off-nominal values for one of the guidance 
parameters with the guidance algorithm being healthy, system does the action 
corresponding to discarding the present computation within $20$ {\tt ms} 
followed by recalculation action within $20$-$40$ {\tt ms}}.
\[
{\tt NGCS\_R5:\ guid\_comp\_error\rightarrow \text{\#\#}1\ act6\ 
\text{\#\#}[1:2]\ act12}
\]

 \item ${\tt NGCS\_R6}$:
{\em As soon as the  system detects error in the execution of vehicle autopilot 
commands, then the system initiates the action to switch to open loop guidance 
within $20$-$40$ {\tt ms}followed by salvage profile initialization action for 
control within $20$-$60$ {\tt ms}. Both the actions are spatially repeated 
twice}.
\[
{\tt NGCS\_R6:\ comm\_exec\_error \rightarrow \text{\#\#}[1:2]\ act1[\sim 2]\ 
\text{\#\#}[1:3]\ act2[\sim 2]}
\]

 \item ${\tt NGCS\_R7}$:
{\em As soon as the system detects any failure in switching Vehicle Autopilot 
digital filters, then the overall action to switch to open loop guidance is 
scheduled within $20$-$60$ {\tt ms} followed by salvage profile initialization 
action for control is scheduled $20$-$40$ {\tt ms}, is repeated 
non-consecutively twice}.
\[
{\tt NGCS\_R7:\ filter\_switch\_fail \rightarrow \text{\#\#}[1:3]\ (act1 
\text{\#\#}[1:2]\ act2)[= 2]}
\]

 \item ${\tt NGCS\_R8}$:
{\em If the system detects non-execution of sequencing(time-based) commands for 
the first time, system will do action corresponding to rescheduling of 
time-based commands within $20$-$60$ {\tt ms} followed by rechecking action to 
check the execution status  within $20$ {\tt ms}. The latter action is repeated 
non-consecutively twice}.
\[
{\tt NGCS\_R8:\ seq\_comm\_non\_exe \rightarrow \text{\#\#}[1:3]\ act7\ 
\text{\#\#}1\ act8[= 2]}
\]

 \item ${\tt NGCS\_R9}$:
{\em As soon as the system detects non-execution of sequencing(time-based) 
commands more than once, system will do action corresponding to switching to 
open loop guidance within $100$ {\tt ms} non-consecutively twice}.
\[
{\tt NGCS\_R9:\ seq\_comm\_non\_exe\ \text{\#\#}[1:\$]\ seq\_comm\_non\_exe 
\rightarrow \text{\#\#}[1:5]\ act1[= 2]}
\]

 \item ${\tt NGCS\_R10}$:
{\em Whenever system detects error in on board arithmetic computation unit, 
system discards all computations within $20$-$40$ {\tt ms} followed by action 
to switch to open loop guidance is executed parallelly in two units within 
$20$-$80${\tt ms} followed by salvage profile initialization action for control 
is executed parallelly in two units within $20$-$60${\tt ms}}.
\[
{\tt NGCS\_R10:\ onboard\_comp\_error \rightarrow \text{\#\#}[1:2]\ act6 
\text{\#\#}[1:4]\ act1[\sim 2] \text{\#\#}[1:3]\ act2[\sim 2]}
\]

 \item ${\tt NGCS\_R11}$:
{\em If the system detects off-normal control parameters with the control 
algorithm being healthy system will trigger action to recalculate guidance 
parameters within $20$-$40$ {\tt ms} followed by triggering action to 
recalculate control parameters within $20$-$60${\tt ms}. The second action is 
repeated in two parallel units}.
\[
{\tt NGCS\_R11:\ abnormal\_control\_param \rightarrow \text{\#\#}[1:3]\ act12\ 
\text{\#\#}[1:3]\ act9[\sim 2]}
\]

 \item ${\tt NGCS\_R12}$:
{\em As soon as the system detects any communication error between navigation, 
guidance, control or sequencing, The system will repeat the overall action to 
switch to open loop guidance within $20$-$80$ {\tt ms} followed by salvage 
profile initialization for control within $20$-$40$ {\tt ms}, non-consecutively 
twice}.
\[
{\tt NGCS\_R12:\ communication\_error \rightarrow\text{\#\#}[1:4]\ (act1 
\text{\#\#}[1:2]\ act2)[= 2]}
\]

 \item ${\tt NGCS\_R13}$:
{\em If the system detects temporary failure in any of the navigation sensor 
channels system will do action to discard sensor data within $20$-$40$ {\tt ms} 
followed by the recalculation of navigation parameters is repeated consecutively 
twice using the redundant channels within $20$-$80${\tt ms}}.
\[
{\tt NGCS\_R13:\ sensor\_ch\_temp\_failure \rightarrow \text{\#\#}[1:2]\ 
act10\ \text{\#\#}[1:4]\ act4[\ast 2]}
\]

 \item ${\tt NGCS\_R14}$:
{\em whenever system has a temporary failure reported in any of the navigation 
sensor channels then system will do rechecking action to check the particular 
channel values for re-induction within $20$-$120$ {\tt ms} and is repeated 
non-consecutively twice}.
\[
{\tt NGCS\_R14:\ sensor\_ch\_temp\_failure \rightarrow \text{\#\#}[1:6]\ act11[= 
2]}
\]

\item ${\tt NGCS\_R15}$:
{\em As soon as the system detects timing failure due to shift in clock 
frequency, system will discards all computations within $20$-$40$ {\tt ms} 
followed by switching to open loop guidance  within $20$ {\tt ms} followed by 
salvage profile initialization for control within $20$-$60$ {\tt ms}. The latter 
two actions are non-consecutively repeated twice}.
\[
{\tt NGCS\_R15:\ clock\_freq\_shift \rightarrow \text{\#\#}[1:2]\ act6\ 
\text{\#\#}1\ (act1 \text{\#\#}[1:3]\ act2)[= 2]}
\]
\end{enumerate}
}

\subsection{Analysis and Discussion}

\begin{table*}[htbp]
\scriptsize
\caption{Possible Options of Action-Events for ${\tt NGCS\_R4}$
\label{table: ngcs_r4}}
\centering
\begin{tabular}{|c||c|c|c|c|c|c|c|c||c|}
\hline
Possible & \multicolumn{8}{|c||}{Action Events (Cycle-wise)} & Computed\\
\cline{2-9}
Options & Cycle-1 & Cycle-2 & Cycle-3 & Cycle-4 & Cycle-5 & Cycle-6 & Cycle-7
& Cycle-8 &  Reliability\\
\hline \hline
\rowcolor{lightgray}
(4A) & ${\tt act6}$ & ${\tt act1}$ & ${\tt act2}$ & & & & & &  $0.9959$\\
\rowcolor{lightgray}
& & & ${\tt act1}$ & ${\tt act2}$ & & & &  &  \\
\hline
\rowcolor{lightgray}
(4B) & ${\tt act6}$ & ${\tt act1}$ & ${\tt act2}$ & & & & & &  $0.9950$\\
\rowcolor{lightgray}
& & & ${\tt act1}$ & & ${\tt act2}$ & & &  &  \\
\hline
\rowcolor{lightgray}
(4C) & ${\tt act6}$ & ${\tt act1}$ & & ${\tt act2}$ & & & & &  $0.9960$\\
\rowcolor{lightgray}
& & & ${\tt act1}$ & ${\tt act2}$ & & & &  &  \\
\hline
\rowcolor{lightgray}
(4D) & ${\tt act6}$ & ${\tt act1}$ & & ${\tt act2}$  & & & & &  $0.9950$\\
\rowcolor{lightgray}
& & & ${\tt act1}$ & & ${\tt act2}$ & & &  &  \\
\hline
\rowcolor{lightgray}
(4E) & & ${\tt act6}$ & ${\tt act1}$ & ${\tt act2}$  & & & & &  $0.9959$\\
\rowcolor{lightgray}
& & & & ${\tt act1}$ & ${\tt act2}$ & & &  &  \\
\hline
\rowcolor{lightgray}
(4F) & & ${\tt act6}$ & ${\tt act1}$ & ${\tt act2}$ & & & & &  $0.9960$\\
\rowcolor{lightgray}
& & & ${\tt act1}$ & & ${\tt act2}$ & & &  &  \\
\hline
\rowcolor{lightgray}
(4G) & & ${\tt act6}$ & ${\tt act1}$ & & ${\tt act2}$  & & & &  $0.9960$\\
\rowcolor{lightgray}
& & & & ${\tt act1}$ & ${\tt act2}$  & & &  &  \\
\hline
\rowcolor{lightgray}
(4H) & & ${\tt act6}$ & ${\tt act1}$ & & ${\tt act2}$   & & & &  $0.9960$\\
\rowcolor{lightgray}
& & & & ${\tt act1}$ & & ${\tt act2}$ & &  &  \\
\hline
\rowcolor{lightgray}
(4I) & ${\tt act6}$ & & ${\tt act1}$ & ${\tt act2}$ & & & & &  $0.9959$\\
\rowcolor{lightgray}
& & & & ${\tt act1}$ & ${\tt act2}$ & & &  &  \\
\hline
\rowcolor{lightgray}
(4J) & ${\tt act6}$ & & ${\tt act1}$ & ${\tt act2}$ & & & & &  $0.9950$\\
\rowcolor{lightgray}
& & & & ${\tt act1}$ & & ${\tt act2}$ & & &    \\
\hline
\rowcolor{lightgray}
(4K) & ${\tt act6}$ & & ${\tt act1}$ & & ${\tt act2}$ & & & &  $0.9960$\\
\rowcolor{lightgray}
& & & & ${\tt act1}$ & ${\tt act2}$ & & & &   \\
\hline
\rowcolor{lightgray}
(4L) & ${\tt act6}$ & & ${\tt act1}$ & & ${\tt act2}$ & & & &  $0.9959$\\
\rowcolor{lightgray}
& & & & ${\tt act1}$ & & ${\tt act2}$ & & &    \\
\hline
\rowcolor{lightgray}
(4M) & & ${\tt act6}$ & & ${\tt act1}$ & ${\tt act2}$  & & & &  $0.9959$\\
\rowcolor{lightgray}
& & & & & ${\tt act1}$ & ${\tt act2}$ & & &  \\
\hline
\rowcolor{lightgray}
(4N) & & ${\tt act6}$ & & ${\tt act1}$ & ${\tt act2}$ & & & & $0.9950$\\
\rowcolor{lightgray}
& & & & & ${\tt act1}$ & & ${\tt act2}$ &  &  \\
\hline
\rowcolor{lightgray}
(4O) & & ${\tt act6}$ & & ${\tt act1}$ & & ${\tt act2}$  & & &  $0.9960$\\
\rowcolor{lightgray}
& & & & & ${\tt act1}$ & ${\tt act2}$  & &  &  \\
\hline
\rowcolor{lightgray}
(4P) & & ${\tt act6}$ & & ${\tt act1}$ & & ${\tt act2}$   & & &  $0.9959$\\
\rowcolor{lightgray}
& & & & & ${\tt act1}$ & & ${\tt act2}$ &  &  \\
\hline
\rowcolor{lightgray}
(4A$\ast$) & ${\tt act6}$ & ${\tt act1}$ & ${\tt act2}$ & & & & & &  $0.9950$\\
\rowcolor{lightgray}
& & & & ${\tt act1}$ & ${\tt act2}$  & & &  &  \\
\hline
\rowcolor{lightgray}
(4B$\ast$) & ${\tt act6}$ & ${\tt act1}$ & ${\tt act2}$ & & & & & &  $0.9950$\\
\rowcolor{lightgray}
& & & & ${\tt act1}$ & & ${\tt act2}$  & &  &  \\
\hline
\rowcolor{lightgray}
(4C$\ast$) & ${\tt act6}$ & ${\tt act1}$ & & ${\tt act2}$ & & & & &  $0.9950$\\
\rowcolor{lightgray}
& & & & ${\tt act1}$ & ${\tt act2}$ & & &  &  \\
\hline
\rowcolor{lightgray}
(4D$\ast$) & ${\tt act6}$ & ${\tt act1}$ & & ${\tt act2}$  & & & & &  $0.9950$\\
\rowcolor{lightgray}
& & & & ${\tt act1}$ & & ${\tt act2}$ & & &    \\
\hline
\rowcolor{lightgray}
(4E$\ast$) & & ${\tt act6}$ & ${\tt act1}$ & ${\tt act2}$  & & & & &  $0.9950$\\
\rowcolor{lightgray}
& & & & & ${\tt act1}$ & ${\tt act2}$ & &  &  \\
\hline
\rowcolor{lightgray}
(4F$\ast$) & & ${\tt act6}$ & ${\tt act1}$ & ${\tt act2}$ & & & & &  $0.9950$\\
\rowcolor{lightgray}
& & & & ${\tt act1}$ & & ${\tt act2}$ & &  &  \\
\hline
\rowcolor{lightgray}
(4G$\ast$) & & ${\tt act6}$ & ${\tt act1}$ & & ${\tt act2}$  & & & &  $0.9950$\\
\rowcolor{lightgray}
& & & & & ${\tt act1}$ & ${\tt act2}$   & &  &  \\
\hline
\rowcolor{lightgray}
(4H$\ast$) & & ${\tt act6}$ & ${\tt act1}$ & & ${\tt act2}$   & & & &  
$0.9950$\\
\rowcolor{lightgray}
& & & & & ${\tt act1}$ & & ${\tt act2}$ &  &  \\
\hline
(4I$\ast$) & ${\tt act6}$ & & ${\tt act1}$ & ${\tt act2}$ & & & & &  $0.9653$\\
& & & & & ${\tt act1}$ & ${\tt act2}$  & &  &  \\
\hline
(4J$\ast$) & ${\tt act6}$ & & ${\tt act1}$ & ${\tt act2}$ & & & & &  $0.9653$\\
& & & & & ${\tt act1}$ & & ${\tt act2}$  & &    \\
\hline
(4K$\ast$) & ${\tt act6}$ & & ${\tt act1}$ & & ${\tt act2}$ & & & &  $0.9653$\\
& & & & & ${\tt act1}$ & ${\tt act2}$ & & &   \\
\hline
(4L$\ast$) & ${\tt act6}$ & & ${\tt act1}$ & & ${\tt act2}$ & & & &  $0.9653$\\
& & & & & ${\tt act1}$ & & ${\tt act2}$ & &    \\
\hline
(4M$\ast$) & & ${\tt act6}$ & & ${\tt act1}$ & ${\tt act2}$  & & & &  $0.9653$\\
& & & & & & ${\tt act1}$ & ${\tt act2}$ & &  \\
\hline
(4N$\ast$) & & ${\tt act6}$ & & ${\tt act1}$ & ${\tt act2}$ & & & & $0.9653$\\
& & & & & & ${\tt act1}$ & & ${\tt act2}$  &  \\
\hline
(4O$\ast$) & & ${\tt act6}$ & & ${\tt act1}$ & & ${\tt act2}$  & & &  $0.9653$\\
& & & & & & ${\tt act1}$ & ${\tt act2}$  &  &  \\
\hline
(4P$\ast$) & & ${\tt act6}$ & & ${\tt act1}$ & & ${\tt act2}$   & & &  
$0.9653$\\
& & & & & & ${\tt act1}$ & & ${\tt act2}$ & \\
\hline
\end{tabular}
\end{table*}

\begin{table*}[htb]
\scriptsize
\caption{Possible Options of Action-Events for ${\tt NGCS\_R13}$
\label{table: ngcs_r13}}
\centering
\begin{tabular}{|c||c|c|c|c|c|c|c||c|}
\hline
Possible & \multicolumn{7}{|c||}{Action Events (Cycle-wise)} & Computed\\
\cline{2-8}
Options & Cycle-1 & Cycle-2 & Cycle-3 & Cycle-4 & Cycle-5 & Cycle-6 & Cycle-7
& Reliability\\
\hline \hline
\rowcolor{lightgray}
(13A) & ${\tt act10}$  & ${\tt act4}$ & ${\tt act4}$ & & & & & $0.9995$\\
\hline
\rowcolor{lightgray}
(13B) & ${\tt act10}$ &  & ${\tt act4}$ & ${\tt act4}$ & & & & $0.9995$\\
\hline
\rowcolor{lightgray}
(13C) & ${\tt act10}$ & & & ${\tt act4}$ & ${\tt act4}$ & & & $0.9995$\\
\hline
(13D)  & ${\tt act10}$ & & & & ${\tt act4}$ & ${\tt act4}$ & &  $0.9790$\\
\hline
\rowcolor{lightgray}
(13E)  & & ${\tt act10}$ & ${\tt act4}$ & ${\tt act4}$  & & & & $0.9995$\\
\hline
\rowcolor{lightgray}
(13F) & & ${\tt act10}$ & & ${\tt act4}$ & ${\tt act4}$ &  & &  $0.9995$\\
\hline
\rowcolor{lightgray}
(13G) & & ${\tt act10}$ & & & ${\tt act4}$ & ${\tt act4}$ & &   $0.9995$\\
\hline
(13H) & & ${\tt act10}$ & & & & ${\tt act4}$ & ${\tt act4}$ &    $0.9790$\\
\hline
\end{tabular}
\end{table*}

Given the reliability specifications, ${\tt NGCS\_R1} - {\tt NGCS\_R15}$ and 
assuming that the sensed-events are present in ${\tt Cycle-0}$, the reliability 
for the correctness specifications, ${\tt NGCS\_C1} - {\tt NGCS\_C15}$ can be 
calculated with respect to each action/control strategy, in the same way as 
shown in Example~\ref{ex: acc_rel}. We present two representative tables, 
namely Table~\ref{table: ngcs_r4} and Table~\ref{table: ngcs_r13}, 
corresponding to the specifications, ${\tt NGCS\_C4}$ and ${\tt NGCS\_C13}$, 
respectively, to demonstrate the result\footnote{For brevity, we are not 
showing the tables for all $15$ specifications mentioned.}. To illustrate the 
computation of reliability values, let us analyze one example action strategy, 
say $(13A)$, from Table~\ref{table: ngcs_r13} which is defined over the 
correctness specification, ${\tt NGCS\_C13}$, and reliability specification, 
${\tt NGCS\_R13}$. Here, the realization of ${\tt NGCS\_C13}$ can happen in {\em 
two} possible ways as per $(13A)$: (i) ${\tt act 10}$ followed by ${\tt act 4}$ 
immediately in the next cycle; or (ii) ${\tt act 10}$ followed by ${\tt act 4}$ 
after a gap of one cycle. Since, it is given that $R_{\tt discard\_sensor\_data} 
= 0.996$ and $R_{\tt recal\_navig\_param}= 0.983$, therefore the computed 
reliability becomes, $R_{{\tt NGCS\_C13}}^{(13A)} = 1 - (1 - R_{\tt 
discard\_sensor\_data} \times R_{\tt recal\_navig\_param})^2 = 0.9995$. The 
\colorbox{lightgray}{highlighted} rows of Table~\ref{table: ngcs_r4}  indicate 
the admissible action-strategies that meet the desired reliability requirements 
for the property, ${\tt NGCS\_C4}$ and ${\tt NGCS\_C13}$, of ${\tt NGC}$ 
subsystem. For example, Table~\ref{table: ngcs_r4} shows that the action 
strategies, $(4A)-(4P)$ and $(4A*)-(4H*)$ are admissible with computed 
reliability values $\geq 0.992$, whereas $(4I*)-(4P*)$ options do not meet the 
desired reliability target.

Our proposed methodology analyzes the given specifications with reliability
targets and searches for appropriate allocation of the action-events
(cycle-wise) so that the minimum number of resources are required to realize
the specifications. Table~\ref{table: ngcs_ro} shows the cycle-wise allocations
for all the action-events (from the reliability specifications of NGC system) 
attributing to the minimum resource requirement, assuming simultaneous 
occurrence of all sensed-events. It may be noted that, here we need a minimum of 
{\em three (3)} resources to realize the given set of specifications, which is 
evident from the last column of Table~\ref{table: ngcs_ro}.

\begin{table*}[htbp]
\scriptsize
\caption{Cycle-wise Action-event Allocation attributing to Minimum Resource 
Requirement (assuming a reliability target of 0.992 for all properties and all 
sensed-events occurring simultaneously)
\label{table: ngcs_ro}}
\centering
\begin{tabular}{|c|c||c|c|c|c|c|c|c|}
\hline
Reliability& Admissible Action & \multicolumn{7}{|c|}{Action-event Allocation 
Schedule}\\
\cline{3-9}
Specification & Strategy Index & Cycle-1 & Cycle-2 & Cycle-3 & Cycle-4 & 
Cycle-5 & Cycle-6 & Cycle-7\\
\hline \hline
${\tt NGCS\_R1}$ & $1G$ & ${\tt act10}$  & ${\tt act1}, {\tt act1}$ &  &  & 
${\tt act2}, {\tt act2}$ &  & \\
\hline
${\tt NGCS\_R2}$ & $2E$ & ${\tt act3}$ &  &  & ${\tt act4}$ & ${\tt act4}$ &  & 
\\
\hline
${\tt NGCS\_R3}$ & $3D$ &  & ${\tt act5}$ &  & ${\tt act4}$ & ${\tt act4}$ &  & 
\\
\hline
${\tt NGCS\_R4}$ & $4A*$ & ${\tt act6}$ & ${\tt act1}$ & ${\tt act1}$ & ${\tt 
act2}$ & ${\tt act2}$ &  & \\
\hline
${\tt NGCS\_R5}$ & $5B$ & ${\tt act6}$ &  & ${\tt act12}$ &  &  &  & \\
\hline
${\tt NGCS\_R6}$ & $6F$ &  & ${\tt act1}, {\tt act1}$ &  &  & ${\tt act2}, {\tt 
act2}$ &  & \\
\hline
${\tt NGCS\_R7}$ & $7D*$ &  & ${\tt act1}$ & ${\tt act1}$ & ${\tt act2}$ & ${\tt 
act2}$ &  & \\
\hline
${\tt NGCS\_R8}$ & $8K$ &  &  & ${\tt act7}$ & ${\tt act8}$ &  &  & ${\tt 
act8}$\\
\hline
${\tt NGCS\_R9}$ & $9G$ &  & ${\tt act1}$ & ${\tt act1}$ &  &  &  & \\
\hline
${\tt NGCS\_R10}$ & $10I$ & ${\tt act6}$ & ${\tt act1}, {\tt act1}$  &  &  & 
${\tt act2}, {\tt act2}$  &  & \\
\hline
${\tt NGCS\_R11}$ & $11I$ &  &  & ${\tt act12}$ &  &  & ${\tt act9}, {\tt act9}$ 
& \\
\hline
${\tt NGCS\_R12}$ & $12D*$ &  & ${\tt act1}$ & ${\tt act1}$ & ${\tt act2}$ & 
${\tt act2}$ &  & \\
\hline
${\tt NGCS\_R13}$ & $13C$ & ${\tt act10}$ &  &  & ${\tt act4}$ & ${\tt act4}$ &  
& \\
\hline
${\tt NGCS\_R14}$ & $14Y$ &  &  &  &  &  & ${\tt act11}$ & ${\tt act11}$\\
\hline
${\tt NGCS\_R15}$ & $15M$ & ${\tt act6}$  & ${\tt act1}$ & ${\tt act1}$ & ${\tt 
act2}$ & ${\tt act2}$ &  & \\
\hline \hline
\multicolumn{2}{|c||}{Action-event Allocation} & ${\tt \langle act3, act6,}$ & 
${\tt \langle act1, act1,}$ & ${\tt \langle act1, act7,}$ & ${\tt \langle act2, 
act4,}$ & ${\tt \langle act2, act2,}$ & ${\tt \langle act9, act9,}$ & ${\tt 
\langle act8,}$\\
\multicolumn{2}{|c||}{(cycle-wise)} & ${\tt act10 \rangle}$ & ${\tt act5 
\rangle}$ & ${\tt act12 \rangle}$ & ${\tt act8 \rangle}$ & ${\tt act4 \rangle}$ 
& ${\tt act11 \rangle}$ & ${\tt act11 \rangle}$\\
\hline
\end{tabular}
\end{table*}

Though the simultaneous occurrence of all sensed-events is the pessimistic-case
and needs to be analyzed for deriving the worst-case resource requirements, but
in practical scenarios, all the sensed-events may not appear simultaneously. If we
assume that the only temporary failures (sensed-events corresponding to
${\tt NGCS\_R2}$, ${\tt NGCS\_R3}$, ${\tt NGCS\_R5}$, ${\tt NGCS\_R8}$,
${\tt NGCS\_R11}$, ${\tt NGCS\_R13}$ and ${\tt NGCS\_R14}$) would happen
simultaneously (failures which  would not lead to the selection of salvage profile)
and the remaining sensed-events (corresponding to ${\tt NGCS\_R1}$,
${\tt NGCS\_R4}$, ${\tt NGCS\_R6}$, ${\tt NGCS\_R7}$, ${\tt NGCS\_R9}$,
${\tt NGCS\_R10}$, ${\tt NGCS\_R12}$ and ${\tt NGCS\_R15}$) occurs $7$ or
more cycles later, the number of resources required is only {\em two (2)} as
shown in Tables~\ref{table: ngcs_ro1} and Table~\ref{table: ngcs_ro2}. This is
the minimum number of resources required, assuming all the sensed-events do
not occur together.

\begin{table*}[htbp]
\small
\caption{Cycle-wise Action-event Allocation attributing to Minimum Resource 
Requirement (assuming a reliability target of 0.992 for all properties and only 
the sensed-events corresponding to ${\tt NGCS\_R2}$, ${\tt NGCS\_R3}$, ${\tt 
NGCS\_R5}$, ${\tt NGCS\_R8}$, ${\tt NGCS\_R11}$, ${\tt NGCS\_R13}$ and ${\tt 
NGCS\_R14}$ occurring simultaneously)
\label{table: ngcs_ro1}}
\centering
\begin{tabular}{|c|c||c|c|c|c|c|c|c|}
\hline
Reliability& Admissible Action & \multicolumn{7}{|c|}{Action-event Allocation 
Schedule}\\
\cline{3-9}
Specification & Strategy Index & Cycle-1 & Cycle-2 & Cycle-3 & Cycle-4 & 
Cycle-5 & Cycle-6 & Cycle-7\\
\hline \hline
${\tt NGCS\_R2}$ & $2D$ &  & ${\tt act3}$ &  & ${\tt act4}$ & ${\tt act4}$ &  & 
\\
\hline
${\tt NGCS\_R3}$ & $3D$ &  & ${\tt act5}$ &  & ${\tt act4}$ & ${\tt act4}$ &  & 
\\
\hline
${\tt NGCS\_R5}$ & $5B$ & ${\tt act6}$ &  & ${\tt act12}$ &  &  &  & \\
\hline
${\tt NGCS\_R8}$ & $8K$ &  &  & ${\tt act7}$ & ${\tt act8}$ &  &  & ${\tt 
act8}$\\
\hline
${\tt NGCS\_R11}$ & $11I$ &  &  & ${\tt act12}$ &  &  & ${\tt act9}, {\tt 
act9}$ & \\
\hline
${\tt NGCS\_R13}$ & $13C$ & ${\tt act10}$ &  &  & ${\tt act4}$ & ${\tt act4}$ & 
 & \\
\hline
${\tt NGCS\_R14}$ & $14W$ &  &  &  &  & ${\tt act11}$ &  & ${\tt act11}$\\
\hline \hline
\multicolumn{2}{|c||}{Action-event Allocation} & ${\tt \langle\ act6,}$ &
${\tt \langle\ act3,}$ & ${\tt \langle\ act7,}$ & ${\tt \langle\ act4,}$ &
${\tt \langle\ act4,}$ & ${\tt \langle\ act9,}$ & ${\tt \langle\ act8,}$\\
\multicolumn{2}{|c||}{(cycle-wise)} & ${\tt act10\ \rangle}$ &
${\tt act5\ \rangle}$ & ${\tt act12\ \rangle}$ & ${\tt act8\ \rangle}$ &
${\tt act11\ \rangle}$ & ${\tt act9\ \rangle}$ & ${\tt act11\ \rangle}$\\
\hline
\end{tabular}
\end{table*}

\begin{table*}[htbp]
\small
\caption{Cycle-wise Action-event Allocation attributing to Minimum Resource 
Requirement (assuming a reliability target of 0.992 for all properties and only 
the sensed-events corresponding to ${\tt NGCS\_R1}$, ${\tt NGCS\_R4}$, ${\tt 
NGCS\_R6}$, ${\tt NGCS\_R7}$, ${\tt NGCS\_R9}$, ${\tt NGCS\_R10}$, ${\tt 
NGCS\_R12}$ and ${\tt NGCS\_R15}$ occurring simultaneously)
\label{table: ngcs_ro2}}
\centering
\begin{tabular}{|c|c||c|c|c|c|c|}
\hline
Reliability& Admissible Action & \multicolumn{5}{|c|}{Action-event Allocation 
Schedule}\\
\cline{3-7}
Specification & Strategy Index & Cycle-1 & Cycle-2 & Cycle-3 & Cycle-4 & 
Cycle-5\\
\hline \hline
${\tt NGCS\_R1}$ & $1G$ & ${\tt act10}$ & ${\tt act1}, {\tt act1}$ &  &  & ${\tt 
act2}, {\tt act2}$\\
\hline
${\tt NGCS\_R4}$ & $4A*$ & ${\tt act6}$ & ${\tt act1}$ & ${\tt act1}$ & ${\tt 
act2}$ & ${\tt act2}$\\
\hline
${\tt NGCS\_R6}$ & $6F$ &  & ${\tt act1}, {\tt act1}$ &  &  & ${\tt act2}, {\tt 
act2}$\\
\hline
${\tt NGCS\_R7}$ & $7D*$ &  & ${\tt act1}$ & ${\tt act1}$ & ${\tt act2}$ & 
${\tt act2}$\\
\hline
${\tt NGCS\_R9}$ & $9G$ & & ${\tt act1}$ & ${\tt act1}$ &  & \\
\hline
${\tt NGCS\_R10}$ & $10I$ & ${\tt act6}$ & ${\tt act1}, {\tt act1}$ &  &  & 
${\tt act2}, {\tt act2}$\\
\hline
${\tt NGCS\_R12}$ & $12D*$ &  & ${\tt act1}$ & ${\tt act1}$ & ${\tt act2}$ & 
${\tt act2}$\\
\hline
${\tt NGCS\_R15}$ & $15M$ & ${\tt act6}$  & ${\tt act1}$ & ${\tt act1}$ & ${\tt 
act2}$ & ${\tt act2}$\\
\hline \hline
\multicolumn{2}{|c||}{Action-event Allocation (cycle-wise)} & ${\tt \langle 
act6, act10 \rangle}$ & ${\tt \langle act1, act1 \rangle}$ & ${\tt \langle 
act1 \rangle}$ & ${\tt \langle act2 \rangle}$ & ${\tt \langle 
act2, act2 \rangle}$\\
\hline
\end{tabular}
\end{table*}

\begin{figure*}[htbp]
\begin{center}
\subfloat[Z3 SMT Solver Response]
 {\includegraphics[scale=0.23]{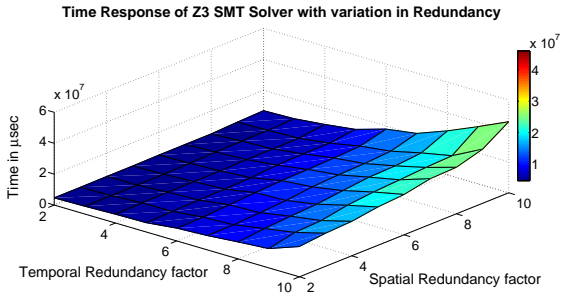}} 
 \quad
\subfloat[CPLEX ILP Solver Response]
 {\includegraphics[scale=0.23]{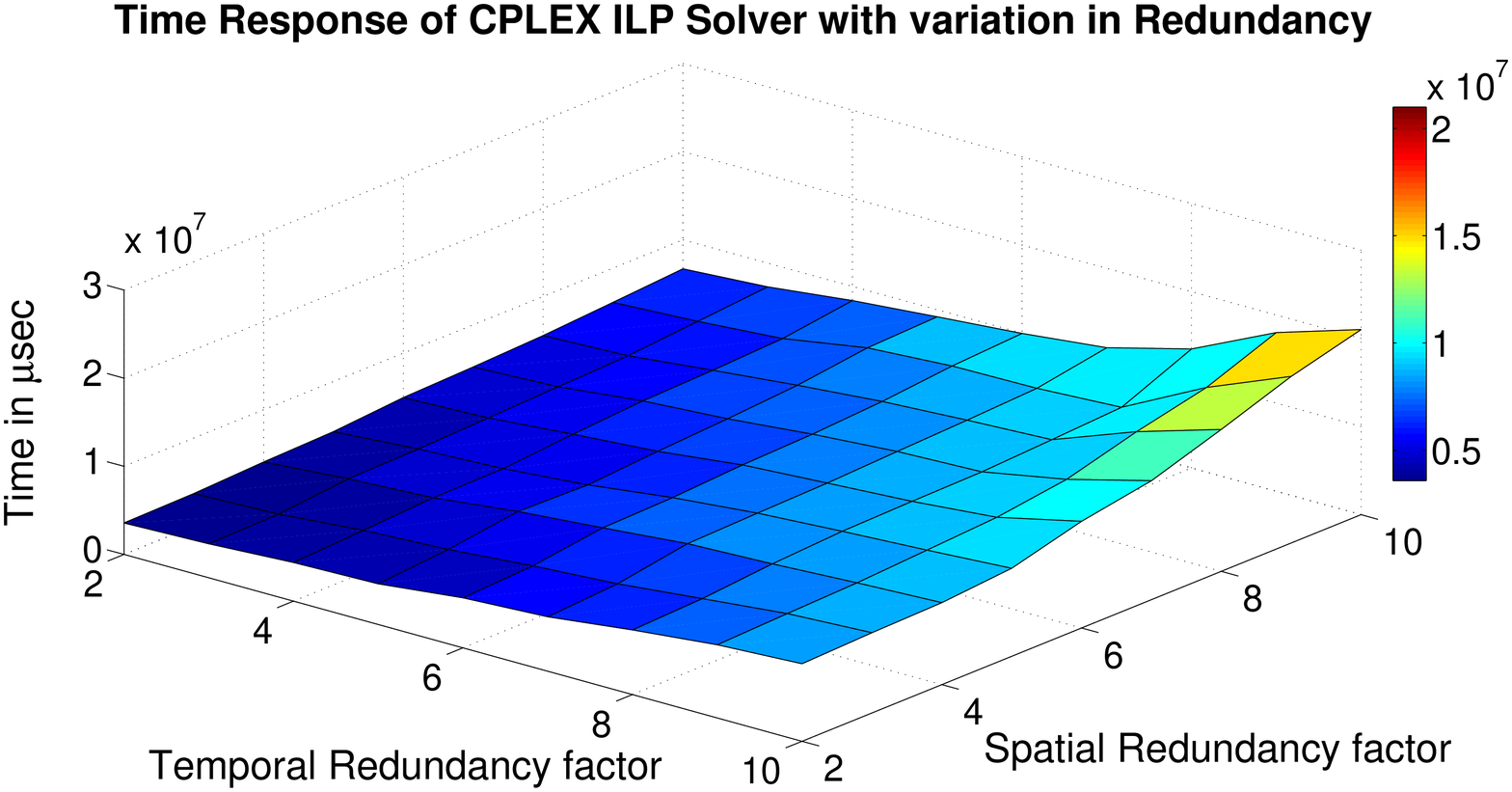}}
\end{center}
\caption{Analysis Time when Redundancy Artifacts Incorporated in Reliability 
Specifications of NGC system}
\label{fig: 3D_graphs1_ngcs}
\end{figure*}

\begin{figure}[htb]
\begin{center}
 {\includegraphics[scale=0.4]{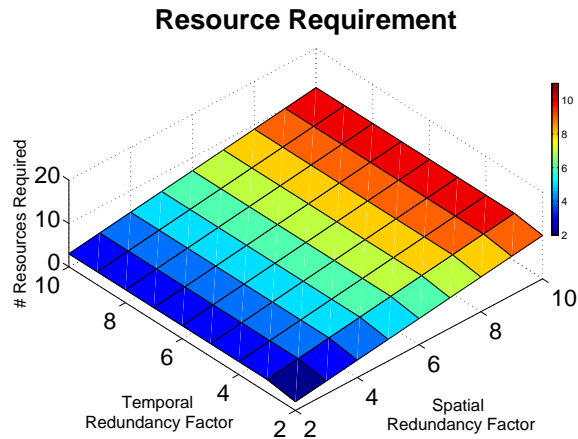}}
\end{center}
\caption{Resource Requirement for Variations in Redundancy Artifacts 
Incorporated in Reliability Specifications of NGC system}
\label{fig: 3D_graphs2_ngcs}
\end{figure}

In addition to this, we also provide Figure~\ref{fig: 3D_graphs1_ngcs} and
Figure~\ref{fig: 3D_graphs2_ngcs} to show how our analysis time and the
minimum resource requirements depend on the increase in temporal and
spatial redundancy factors to improve reliability.
In Figure~\ref{fig: 3D_graphs1_ngcs} the response of Z3 and CPLEX solvers to 
variations in number of specifications is shown. The Z-axis in the graphs 
represents the time required to find the minimum number of resources in 
Microseconds. Where as in Figure~\ref{fig: 3D_graphs2_ngcs}, the Z-axis 
indicates the resource requirement predicted by the algorithm. The X-axis 
indicates the spatial redundancy artifact incorporated in the reliability 
specifications (varied from $2$ to $10$) in all the graphs. The Y-axis indicates
the temporal redundancy artifact incorporated in the reliability specifications 
(varied from $2$ to $10$) in all the graphs. As the redundancy incorporated 
increases, the execution time varies exponentially. Here the increase along the 
X-axis and Y-axis results in increase in number of timing constraints. This
increase in number of timing constraints can be attributed to increase in number
of action-events due to increased redundancy. There will be variation in number
of resource constraints, implied by increase in number of action-events.

Comparing the performance of the Z3 solver with the CPLEX solver, the CPLEX is 
significantly faster in the above scenario. The CPLEX ILP solver being a COP 
solver, its search space is limited to the optimal hyper plane which maximizes 
the objective function making it more efficient. The resource requirement varies 
linearly as redundancy incorporated increases. The resource required is a linear 
function of incorporated redundancy.

\section{Conclusion} \label{sec: conclusion}
Early-stage estimation of required resources from formal reliability
specifications could be helpful in reducing the cost of a safety-critical
system. This is typically decided at the design time of such systems and often
the choice remain pessimistic by putting additional resources than required. In
this work, we propose an algorithm to find the optimum number of processing
resources required to guarantee the reliability target of given specifications.
The problem is modeled as a {\em Constraint Satisfaction Problem (CSP)} or {\em
Constraint Optimization Problem (COP)} which can be solved efficiently using an
SMT solver or an ILP solver, respectively. Such an analysis is really helpful
in safety-critical embedded system development. The practicality of the
proposed method has been shown over the case-studies from automotive as well as
avionics domain. Besides, the scalability experiments reveal the applicability
and efficacy of the proposed method in complex systems.

The work presented here is an enabler for early prediction of computing
resources in embedded CPS control systems based on their reliability
requirements. However, it is pertinent to note that this work considers the
sense, action and outcome of events from an abstract (specification) level of
system design, and therefore we restrict ourselves without considering the
communication delays between components and other details from this early
design stages. Moreover, scheduling provisions of the action events is also
kept as static in order to accurately specify the temporal boundaries in the
specification. In future, we plan to extend the specification and analysis
framework to incorporate more realistic communication delays and may also keep
provisions for dynamic scheduling of the events. Such specification-level
extensions will require much improved analysis involving the interference in
component-level activities as well. In addition to this, along with the
unreliable computational platform producing outcomes failures, ramifications
due to the unreliable nature of sensed and action event generation also needs to
be explored in future for completeness.

\subsection*{Acknowledgement}
The authors thank Noel Philip (Quality Engineer, Systems Reliability Entity, 
VSSC-ISRO) for helping in developing the NGC subsystem case-study. Besides, P.P.
Chakrabarti acknowledges the partial support from DST, India and P. Dasgupta 
acknowledges the partial support from FMSAFE project (under IMPRINT-India).

\newpage

\bibliographystyle{plain}
\bibliography{reference}

\begin{thebibliography}{10}

\bibitem{AD1994TCS}
R.~Alur and D.~L. Dill.
\newblock {A Theory of Timed Automata}.
\newblock {\em {Theoretical Computer Science}}, {126}({2}):{183--235}, {1994}.

\bibitem{AIKY1992CAV}
R.~Alur, A.~Itai, R.~P. Kurshan, and M.~Yannakakis.
\newblock {Timing Verification by Successive Approximation}.
\newblock In {\em {Computer-Aided Verification (CAV)}}, pages {137--150},
  {1992}.

\bibitem{AK1984C}
A.~Avizienis and J.~P.~J. Kelly.
\newblock {Fault Tolerance by Design Diversity: Concepts and Experiments}.
\newblock {\em {IEEE Computers}}, {17}({8}):{67--80}, {1984}.

\bibitem{BFMSPP2003CASES}
M.~Baleani, A.~Ferrari, L.~Mangeruca, A.~L. Sangiovanni-Vincentelli, M.~Peri,
  and S.~Pezzini.
\newblock {Fault-tolerant Platforms for Automotive Safety-critical
  Applications}.
\newblock In {\em {Proceedings of the International Conference on Compilers,
  Architecture and Synthesis for Embedded Systems (CASES)}}, pages {170--177},
  {2003}.

\bibitem{BAR2009SAT}
C.~W. Barrett, R.~Sebastiani, S.~A. Seshia, and C.~Tinelli.
\newblock {Satisfiability Modulo Theories}.
\newblock {\em {Handbook of satisfiability}}, {185}:{825--885}, {2009}.

\bibitem{BCCZ1999LNCS}
A.~Biere, A.~Cimatti, E.~M. Clarke, and Y.~Zhu.
\newblock {Symbolic Model Checking without BDDs}.
\newblock {\em {Lecture Notes in Computer Science}}, {1579}:{193--207}, {1999}.

\bibitem{CACGWZ2016DT}
S.~Chakraborty, Md.~A. Al-Faruque, W.~Chang, D.~Goswami, M.~Wolf, and Q.~Zhu.
\newblock {Automotive Cyber-Physical Systems: A Tutorial Introduction}.
\newblock {\em {IEEE Design \& Test}}, {33}({4}):{92--108}, {Aug.} {2016}.

\bibitem{CBRZ2001FMSD}
E.~M. Clake, A.~Biere, R.~Raimi, and Y.~Zhu.
\newblock {Bounded Model Checking Using Satisfiability Solving}.
\newblock {\em {The Journal of Formal Methods in System Design}},
  {19}({1}):{7--34}, {2001}.

\bibitem{CGP2000BOOK}
E.~M. Clarke, O.~Grumberg, and D.~A. Peled.
\newblock {\em {Model Checking}}.
\newblock {MIT Press}, {2000}.

\bibitem{DDD1998JPDC}
D.~Das, P.~Dasgupta, and P.~P. Das.
\newblock {A Heuristic for the Maximum Processor Requirement for Scheduling
  Layered Task Graphs with Cloning}.
\newblock {\em {Journal of Parallel and Distributed Computing}},
  {49}({2}):{169--181}, {1998}.

\bibitem{D2006BOOK}
P.~Dasgupta.
\newblock {\em {A Roadmap for Formal Property Verification}}.
\newblock {Springer}, {2006}.

\bibitem{DDR2010DATE}
M.~G. Dixit, P.~Dasgupta, and S.~Ramesh.
\newblock {Taming the Component Timing: A CBD Methodology for Real-time
  Embedded Systems}.
\newblock In {\em {Design Automation and Test in Europe (DATE)}}, pages
  {1649--1652}, {2010}.

\bibitem{DRD2014FAC}
M.~G. Dixit, S.~Ramesh, and P.~Dasgupta.
\newblock {Time-budgeting: A Component Based Development Methodology for
  Real-time Embedded Systems}.
\newblock {\em {Formal Aspects of Computing Journal}}, {26}({3}):{591--621},
  {2014}.

\bibitem{GRW1988TR}
R.~Geist, R.~Raynolds, and J.~Westall.
\newblock {Selection of a Checkpoint Interval in a Critical-Task Environment}.
\newblock {\em {IEEE Transactions on Reliability}}, {37}({4}):{395--400},
  {1988}.

\bibitem{GMBSC2014ISIC}
D.~Goswami, D.~M{\"u}ller-Gritschneder, T.~Basten, U.~Schlichtmann, and
  S.~Chakraborty.
\newblock {Fault-tolerant Embedded Control Systems for Unreliable Hardware}.
\newblock In {\em {2014 International Symposium on Integrated Circuits
  (ISIC)}}, pages {464--467}, {2014}.

\bibitem{HDC2016JAL}
A.~Hazra, P.~Dasgupta, and P.~P. Chakrabarti.
\newblock {Formal Assessment of Reliability Specifications in Embedded
  Cyber-Physical Systems}.
\newblock {\em {Elsevier Journal of Applied Logic (JAL)}}, {18}:{71--104},
  {Nov.} {2016}.

\bibitem{HGDP2013TVLSI}
A.~Hazra, S.~Goyal, P.~Dasgupta, and A.~Pal.
\newblock {Formal Verification of Architectural Power Intent}.
\newblock {\em {IEEE Transactions on Very Large Scale Integration Systems
  (TVLSI)}}, {21}({1}):{78--91}, {January} {2013}.

\bibitem{HMDPHBM2013TCAD}
A.~Hazra, R.~Mukherjee, P.~Dasgupta, A.~Pal, K.~M. Harer, A.~Banerjee, and
  S.Mukherjee.
\newblock {POWER-TRUCTOR: An Integrated Tool Flow for Formal Verification and
  Coverage of Architectural Power Intent}.
\newblock {\em {IEEE Transactions on Computer-Aided Design of Integrated
  Circuits and Systems (TCAD)}}, {32}({11}):{1801--1813}, {November} {2013}.

\bibitem{CPLEX}
{IBM}.
\newblock {CPLEX Optimizer}.
\newblock {\url{https://www.ibm.com/analytics/cplex-optimizer}}.

\bibitem{J1989BOOK}
B.~Johnson.
\newblock {\em {Design and Analysis of Fault Tolerant Digital Systems}}.
\newblock {Addison Wesley}, {MA}, {1989}.

\bibitem{KH1980TR}
M.~Kameyama and T.~Higuchi.
\newblock {Design of Dependent Failure-tolerant Microprocessor System using
  Triple Modular Redundancy}.
\newblock {\em {IEEE Transactions on Reliability}}, {C-29}({2}):{202--206},
  {1980}.

\bibitem{KRH2017DAC}
P.~Khanna, C.~Rebeiro, and A.~Hazra.
\newblock {XFC: A Framework for Exploitable Fault Characterization in Block
  Ciphers}.
\newblock In {\em {the Proceeding of 54th Design Automation Conference (DAC)}},
  pages {8:1--8:6}, {2017}.

\bibitem{K1999RTCSA}
B.~K. Kim.
\newblock {Reliability Analysis of Real-time Controllers with Dual-modular
  Temporal Redundancy}.
\newblock In {\em {Sixth International Conference on Real-Time Computing
  Systems and Applications}}, pages {364--371}, {1999}.

\bibitem{KS1996TC}
H.~Kim and K.~G. Shin.
\newblock {Design and Analysis of an Optimal Instruction Retry Policy for TMR
  Controller Computers}.
\newblock {\em {IEEE Transactions on Computers}}, {45}({11}):{1217--1225},
  {1996}.

\bibitem{KK2004JRTS}
J.~K. Kim and B.~K. Kim.
\newblock {Probabilistic Schedulability Analysis of Harmonic Multi-Task Systems
  with Dual-Modular Temporal Redundancy}.
\newblock {\em {Journal of Real-Time System}}, {26}({2}):{199--222}, {2004}.

\bibitem{KS1993TR}
C.~M. Krishna and A.~D. Singh.
\newblock {Reliability of Checkpointed Real-Time Systems using Time
  Redundancy}.
\newblock {\em {IEEE Transactions on Reliability}}, {42}({3}):{427--435},
  {1993}.

\bibitem{L2005BOOK}
W.~K. Lam.
\newblock {\em {Hardware Design Verification: Simulation and Formal
  Method-Based Approaches}}.
\newblock {Prentice Hall Modern Semiconductor Design Series}, {2005}.

\bibitem{L2008ISORC}
E.~A. Lee.
\newblock {Cyber Physical Systems: Design Challenges}.
\newblock In {\em {11th IEEE International Symposium on Object and
  Component-Oriented Real-Time Distributed Computing (ISORC)}}, pages
  {363--369}, {2008}.

\bibitem{LG2011AIDAC}
R.~M. Lima and I.~E. Grossmann.
\newblock {Computational Advances in Solving Mixed Integer Linear Programming
  Problems}.
\newblock In {\em {AIDAC}}, pages {151--160}, {2011}.

\bibitem{LA1978FTCS}
C.~Liming and A.~Avizienis.
\newblock {N-Version Programming: A Fault-Tolerance Approach to Reliability}.
\newblock In {\em {Proceedings of Fault-Tolerant Computing Symposium}}, pages
  {113--119}, {1978}.

\bibitem{LMM2002EJOR}
A.~Lodi, S.~Martello, and M.~Monaci.
\newblock {Two-dimensional packing problems: A survey}.
\newblock {\em {European Journal of Operational Research}},
  {141}({2}):{241--252}, {2002}.

\bibitem{LCE1989FTCS}
P.~R. Lorczak, A.~K. Caglayan, and D.~E. Eckhardt.
\newblock {A Theoretical Investigation of Generalized Voters for Redundant
  Systems}.
\newblock In {\em {Proceedings of Fault-Tolerant Computing Symposium}}, pages
  {444--451}, {1989}.

\bibitem{M2014LNCS}
O.~Maler.
\newblock {The Unmet Challenge of Times Systems}.
\newblock {\em {Lecture Notes in Computer Science}}, {8415}:{177--192}, {2014}.

\bibitem{M1992PHDTHESIS}
K.~L. McMillan.
\newblock {\em {Symbolic Model Checking -- An Approach to the State Explosion
  Problem}}.
\newblock PhD thesis, {Carnegie Mellon University}, {May} {1992}.

\bibitem{MT2012EUROSTAT}
B.~Meindl and M.~Templ.
\newblock {Analysis of Commercial and Free and Open Source Solvers for Linear
  Optimization Problems}.
\newblock In {\em {Eurostat and Statistics Netherlands within the project
  ESSnet on common tools and harmonised methodology for SDC in the ESS}},
  {2012}.

\bibitem{DB2008TACAS}
L.~De Moura and N.~Bj{\o}rner.
\newblock {Z3: An Efficient SMT Solver}.
\newblock In {\em {Proceedings of Theory and Practice of Software (ETAPS) and
  International Conf. on Tools and Algorithms for Construction and Analysis of
  Systems (TACAS)}}, pages {337--340}, {2008}.

\bibitem{PLW1996TPDS}
M.~A. Palis, J.-C. Liou, and D.~S.~L. Wei.
\newblock {Task Clustering and Scheduling for Distributed Memory Parallel
  Architectures}.
\newblock {\em {IEEE Transactions on Parallel and Distributed Systems}},
  {7}({1}):{46--55}, {Jan.} {1996}.

\bibitem{PCS2008TCAD}
C.~Pinello, L.~P. Carloni, and A.~L. Sangiovanni-Vincentelli.
\newblock {Fault-Tolerant Distributed Deployment of Embedded Control Software}.
\newblock {\em {IEEE Transactions on Computer-Aided Design of Integrated
  Circuits and Systems (TCAD)}}, {27}({5}):{906--919}, {May} {2008}.

\bibitem{SK1994TC}
K.~G. Shin and H.~Kim.
\newblock {A Time Redundancy Approach to TMR Failures using Fault-state
  Likelihoods}.
\newblock {\em {IEEE Transactions on Computers}}, {43}({10}):{1151--1162},
  {1994}.

\bibitem{SS1992BOOK}
D.~P. Siewiorek and R.~Swarz.
\newblock {\em {Reliable Computer Systems: Design and Evaluation}}.
\newblock {Digital Press}, {Burlingtom, MA}, {1992}.

\bibitem{SS2015BOOK}
B.~N. Suresh and K.~Sivan.
\newblock {\em {Integrated Design for Space Transportation System}}.
\newblock {Springer}, {2015}.

\bibitem{system_verilog}
{SystemVerilog LRM}.
\newblock {\em SystemVerilog LRM 3.1a by Accellera}.
\newblock {www.systemverilog.org}, {2004}.

\end{thebibliography}

\end{document}